%% file: main.tex
\pdfoutput=1
\RequirePackage[T1]{fontenc}
\documentclass[12pt]{article}
\usepackage{lmodern}
\usepackage{amsmath}
\usepackage{amssymb}
\usepackage{amsfonts}
\usepackage{bm}
\usepackage{graphicx}
\usepackage{systeme}
\usepackage{enumerate}
\usepackage{comment}
\usepackage{here}
\usepackage{mathrsfs}
\usepackage{url}
\usepackage{braket}
\usepackage{cite}
\usepackage[svgnames]{xcolor}
\usepackage[colorlinks,citecolor=DarkGreen,linkcolor=FireBrick,urlcolor=FireBrick,breaklinks=true,linktoc=all]{hyperref}
\usepackage{setspace} 
\usepackage[top=30truemm,bottom=30truemm,left=25truemm,right=25truemm]{geometry}
\usepackage{authblk}

\title{Foliated-Exotic Duality in Fractonic $BF$ Theories}

\author{Kantaro Ohmori}
\author{Shutaro Shimamura}
\affil{\textit{Department of Physics, The University of Tokyo, Bunkyo-ku, Tokyo 113-0033, Japan}}

\date{}

\renewcommand{\L}{\mathcal{L}}
\newcommand{\Z}{\mathbb{Z}}

\usepackage{xcolor}

\usepackage[status=draft,inline,nomargin]{fixme}
\fxusetheme{color}
\FXRegisterAuthor{ko}{ako}{KO}

\FXRegisterAuthor{shu}{ashu}{\color{magenta}SHU}
    
\begin{document}
\maketitle

\vspace{1.5cm}

\begin{abstract}
    There has been proposed two continuum descriptions of fracton systems: foliated quantum field theories (FQFTs) and exotic quantum field theories. 
    Certain fracton systems are believed to admit descriptions by both, and hence a duality is expected between such a class of FQFTs and exotic QFTs.
    In this paper we study this duality in detail for concrete examples in $2+1$ and $3+1$ dimensions. 
    In the examples, both sides of the continuum theories are of $BF$-type, and we find the explicit correspondences of gauge-invariant operators, gauge fields, parameters, and allowed singularities and discontinuities. 
    This deepens the understanding of dualities in fractonic quantum field theories.
\end{abstract}

\thispagestyle{empty}
\clearpage
\addtocounter{page}{-1}

\newpage

\setlength{\parskip}{2mm}
\setlength{\abovedisplayskip}{10pt} 
\setlength{\belowdisplayskip}{10pt} 
\numberwithin{equation}{section}

\tableofcontents

\newpage

\input{intro}

\newpage

\input{BF-type_21}

\input{BF-type_31}

\input{conclusion}

\subsection*{Acknowledgement}
We thank Yutaka Matsuo and Go Noshita for their contributions in the discussions made in the early stage of the work. We also thank Pranay Gorantla for his helpful comments on the draft.
KO is supported in part by JSPS KAKENHI Grant-in-Aid, No.22K13969 and the Simons Collaboration on Global Categorical Symmetries.
SS is supported by the World-leading INnovative Graduate Study Program for Frontiers of Mathematical Sciences and Physics, The University of Tokyo.

\input{appendixa}

\bibliography{ref} 
\bibliographystyle{utphys}

\end{document}

%% file: intro.tex
\section{Introduction}

A Fracton phase is a new kind of phase of matter that exhibits excitations with restricted mobility, which can only move in certain dimensional submanifolds (see \cite{Nandkishore:2018sel,Pretko:2020cko} for reviews).
The characteristic excitations are called fractons, lineons, and planons, depending on the spatial dimension of the excitation. %
Such fracton models, studied as lattice models in condensed matter physics \cite{Haah:2011drr,Vijay:2015mka,Vijay:2016phm,Shirley:2017suz}, have various novel properties: a new type of symmetry and the exponential growth of ground state degeneracy in terms of the linear sizes of the system. %
The fracton systems are not only theoretically interesting in its own right, but expected to be applied to quantum information \cite{Haah:2011drr,Brown:2019hxw,Khemani:2019vor} and gravity \cite{Pretko:2017fbf}.

While fracton phases first appeared in lattice systems,
one would also expect a continuum description in the low-energy limit of a lattice system.
There have been proposed such descriptions by continuum quantum field theories (QFTs) in various situations\cite{Pretko:2018jbi,Slagle:2017wrc,Seiberg:2019vrp,Seiberg:2020bhn,Seiberg:2020wsg,Seiberg:2020cxy,Gorantla:2020xap,Gorantla:2021bda,Burnell:2021reh,Geng:2021cmq,Slagle:2018swq,Slagle:2020ugk,Hsin:2021mjn,You:2019ciz,Yamaguchi:2021qrx,Yamaguchi:2021xeq,Katsura:2022xkg}. %
The QFTs do not have the Lorentz invariance or even the full rotational invariance, and can have the discontinuous field configurations. In the low-energy descriptions, the gapped %
excitations are not dynamical and arise as the gauge-invariant defects. 
The identification and construction of these QFTs are based on the subsystem symmetry, which is one of the generalizations of symmetry. %
A subsystem symmetry is a symmetry that acts on a spatial submanifold, e.g.\ a plane along a particular directions, and can have different values on each submanifold \cite{Seiberg:2019vrp}.\footnote{While a subsystem symmetry is similar to a higher form symmetry\cite{Gaiotto:2014kfa} as its corresponding symmetry operator has codimension higher than one, the operator is not topological in the directions out of the submanifold.} %

For lattice models, some fracton models can be written as foliated fracton phases \cite{Shirley:2017suz,Shirley:2018nhn,Shirley:2018vtc,Shirley:2022wri}. A foliation is a decomposition of a manifold and regarding it as a stack of an infinite number of submanifolds. For example, the X-cube model \cite{Vijay:2016phm}, which is a gapped fracton lattice model in $3+1$ dimensions, can be written as a stack of the $(2+1)$-dimensional toric codes \cite{Kitaev:1997wr} by using foliations \cite{Shirley:2017suz}. For QFTs, there are fractonic QFTs coupled to %
foliations, which are called foliated quantum filed theories (FQFTs) \cite{Slagle:2018swq,Slagle:2020ugk,Hsin:2021mjn}. 
On the other hand, some fractonic QFTs can be written as tensor gauge theories \cite{Pretko:2016kxt,Pretko:2018jbi,Seiberg:2020bhn,Seiberg:2020wsg,Seiberg:2020cxy,Gorantla:2020xap} respecting the lattice rotational symmetries, which we call the exotic QFTs \cite{Geng:2021cmq}. The continuum QFT description of the X-cube model can be written as $BF$-type theories in terms of both a foliated QFT in the flat foliations\footnote{The foliation is characterized by a foliation filed $e$. The foliation is flat when $e$ is flat, i.e., $d e=0$. See Section \ref{ffield}.} and an exotic QFT \cite{Slagle:2017wrc,Seiberg:2020cxy}. 
The foliated and exotic descriptions are believed to represent the same physics, but the duality between them has not been made clear. %

In this paper, we will consider the foliated and exotic $BF$-type theories in $2+1$ and $3+1$ dimensions. In $2+1$ dimensions, the $BF$-type theories are the continuum description of the $\Z_N$ plaquette Ising model (see \cite{Johnston:2016mbz} for a review) and the $\Z_N$ lattice tensor gauge theory \cite{Seiberg:2020bhn}. In $3+1$ dimensions, the $BF$-type theories are the continuum description of the X-cube model and the $\Z_N$ lattice tensor gauge theory \cite{Seiberg:2020cxy}. 

The goal of this paper is to show the explicit correspondences of the gauge fields and parameters between the foliated $BF$ theory and the exotic $BF$ theory, completing the previous observation made in \cite{Hsin:2021mjn}. We will see that both foliated and exotic $BF$ theories have the same type of gauge-invariant operators and subsystem symmetries, and by matching the operators, we will derive the correspondences of the fields and parameters. %
It is novel to exhibit the explicit correspondences between the foliated fields, including the bulk fields, and the exotic tensor gauge fields.
This establishes the duality between the foliated and exotic $BF$ theories, which we call the foliated-exotic duality.

The organization of the rest of the paper is as follows.
In Section \ref{section 21}, we will discuss the $BF$-type theories in $2+1$ dimensions. In Section \ref{2+1 fBF}, we will consider the foliated $BF$ theory with attention to singularities and discontinuities. The foliated $BF$ Lagrangian in the flat foliations\footnote{The superscripts $k$ index the directions of the foliations. The subscripts in Table~\ref{2corresptable},~\ref{3corresptable}~and~\ref{3corresptable2} are the spatial indices.} is
\begin{align}
    \L_f = \sum^{2}_{k=1} \frac{iN}{2\pi} (dB^k + b)\wedge A^k \wedge dx^k + \frac{iN}{2\pi} b \wedge da\,. 
\end{align}
In Section \ref{2+1 eBF}, we will review the exotic $BF$ theory \cite{Seiberg:2020bhn}. The exotic $BF$ Lagrangian\footnote{In the exotic theories, the superscripts and subscripts are the spacetime indices. As the metric is flat, we do not need to distinguish them.} is
\begin{align}
    \L_e = \frac{i N}{2\pi}\phi^{12}(\partial_0 A_{12} - \partial_{1}\partial_{2} A_0) \,. 
\end{align}
In Section \ref{2+1 corr}, we show the explicit correspondences between them by matching the gauge-invariant operators. In order to match the gauge-invariant operators, we need to modify the strip operators in the foliated $BF$ theory. The modification turns out to be only by an operator that is not remotely detectable \cite{Kong:2014qka,Johnson-Freyd:2020usu}. %
The correspondences of the gauge fields and parameters are shown in Table \ref{2corresptable}. In the correspondences, the singularities and discontinuities are also matched.
In Section \ref{section31}, we will discuss $BF$-type theories in $3+1$ dimensions as in the case of $2+1$ dimensions. In Section \ref{31fbf}, we will review the foliated $BF$ theory \cite{Slagle:2018swq,Slagle:2020ugk,Hsin:2021mjn}. The foliated $BF$ Lagrangian in the flat foliations is
\begin{align}
    \L_f = \sum^{3}_{k=1} \frac{iN}{2\pi} (dB^k + b)\wedge A^k \wedge dx^k + \frac{iN}{2\pi} b \wedge da\,. 
\end{align}
In Section \ref{31ebf}, we will review the exotic $BF$ theory \cite{Slagle:2017wrc,Seiberg:2020cxy}. The exotic $BF$ Lagrangian is
\begin{align}
    \L_e = \frac{i N}{2\pi}\sum_{i,j} \left( \frac{1}{2} A_{ij} (\partial_0 \hat{A}^{ij} - \partial_k \hat{A}^{k(ij)}_0) + \frac{1}{2} A_0 \, \partial_i \partial_j \hat{A}^{ij} \right) \,. 
\end{align}
In Section \ref{31corr}, we show the explicit correspondences between them. The correspondences of the gauge fields and parameters are shown in Table \ref{3corresptable}, \ref{3corresptable2}. In Appendix \ref{appendixa}, we will consider the electric-magnetic dual descriptions of the $BF$-type theories in $2+1$ dimensions.

Along the way, we find that there are gauge-invariant operators that cannot be remotely detected by other spatially placed operators, but represents a time-like symmetry \cite{Gorantla:2022eem}. %
This makes a contrast to the case of ordinary topological order or topological field theory, where every operator is remotely detectable. %

The establishment of the foliated-exotic duality deepens the understanding of both of the continuum descriptions of the fractonic systems. In general it is not known when a fractonic system admits a description by a foliated or an exotic QFT, and this result will be a clue in this interesting question. It would also serve as a starting point of exploring more general dualities in quantum field theories without Lorentz invariance.

\begin{table}[htbp]
\centering
\begin{align*}
\left. \begin{array}{|c|c|c|c|}
\hline 
\multicolumn{2}{|c|}{} & \multicolumn{2}{|c|}{}\\
\multicolumn{2}{|c|}{\text{The foliated $BF$ theory}} 
&
\multicolumn{2}{|c|}{\text{The exotic $BF$ theory}} \\
\multicolumn{2}{|c|}{} & \multicolumn{2}{|c|}{} \\
\hline
&&& \\
 \text{Gauge fields} & \text{Gauge} & \text{Gauge fields} & \text{Gauge} \\
 \text{and parameters} & \text{transformations} & \text{and parameters} & \text{transformations} \\
&&& \\
\hline 
&&&\\
a_0  & \partial_0 \lambda & A_0 & \partial_0 \alpha \\
&&&\\
\hline 
&&&\\
A^k_0 + \partial_0 a_k  & \partial_0 \partial_k \lambda & \partial_k A_0 & \partial_k \partial_0 \alpha \\
(k=1,2) & (k=1,2) & (k=1,2) & (k=1,2) \\
&&&\\
\hline 
&&&\\
A^k_i + \partial_i a_k  & \partial_i \partial_k \lambda & A_{12} & \partial_1 \partial_2 \alpha \\
((k,i)=(1,2),(2,1)) & ((k,i)=(1,2),(2,1)) & &  \\
&&&\\
\hline 
&&&\\
\lambda  & 2\pi \xi^1 + 2\pi \xi^2 & \alpha & 2\pi \tilde{n}^1 + 2\pi \tilde{n}^2 \\
&&&\\
\hline 
&&&\\
\xi^k  &  & \tilde{n}^k &  \\
(k=1,2) & & (k=1,2) &  \\
&&&\\
\hline 
&&&\\
B^1 - B^2  & 2\pi m^1 - 2\pi m^2 & \phi^{12} & 2\pi \tilde{m}^1 - 2\pi \tilde{m}^2 \\
&&&\\
\hline 
&&&\\
m^k  &  &  \tilde{m}^k &  \\
(k=1,2) & & (k=1,2) &  \\
&&&\\
\hline \end{array}\right. 
\end{align*}
\caption{The correspondences of the gauge fields and parameters between the foliated $BF$ theory and the exotic $BF$ theory in $2+1$ dimensions.}
\label{2corresptable}
\end{table}

\begin{table}[htbp]
\centering
\begin{align*}
\left. \begin{array}{|c|c|c|c|}
\hline 
\multicolumn{2}{|c|}{} & \multicolumn{2}{|c|}{}\\
\multicolumn{2}{|c|}{\text{The foliated $BF$ theory}} 
&
\multicolumn{2}{|c|}{\text{The exotic $BF$ theory}} \\
\multicolumn{2}{|c|}{} & \multicolumn{2}{|c|}{} \\
\hline
&&& \\
 \text{Gauge fields} & \text{Gauge} & \text{Gauge fields} & \text{Gauge} \\
 \text{and parameters} & \text{transformations} & \text{and parameters} & \text{transformations} \\
&&& \\
\hline 
&&&\\
a_0  & \partial_0 \lambda & A_0 & \partial_0 \alpha \\
&&&\\
\hline 
&&&\\
A^k_0 + \partial_0 a_k  & \partial_0 \partial_k \lambda & \partial_k A_0 & \partial_k \partial_0 \alpha \\
(k=1,2,3) & (k=1,2,3) & (k=1,2,3) & (k=1,2,3) \\
&&&\\
\hline 
&&&\\
A^k_i + \partial_i a_k  & \partial_i \partial_k \lambda & A_{ki} & \partial_k \partial_i \alpha \\
(k\neq i \,,   & (k\neq i \,,  & (k\neq i \,,  & (k\neq i \,,  \\
k,i \in \{1,2,3\}) & k,i \in \{1,2,3\}) &  k,i \in \{1,2,3\}) & k,i \in \{1,2,3\}) \\
&&&\\
\hline 
&&&\\
\lambda  & 2\pi \xi^1 + 2\pi \xi^2 + 2\pi \xi^3 & \alpha & 2\pi \tilde{n}^1 + 2\pi \tilde{n}^2 +2\pi \tilde{n}^3 \\
&&&\\
\hline 
&&&\\
\xi^k  &  & \tilde{n}^k &  \\
(k=1,2,3) & & (k=1,2,3) &  \\
&&&\\
\hline \end{array}\right. 
\end{align*}
\caption{The correspondences of the gauge fields and parameters between the foliated $BF$ theory and the exotic $BF$ theory in $3+1$ dimensions (the $A$-type and bulk fields).}
\label{3corresptable}
\end{table}

\begin{table}[htbp]
\centering
\begin{align*}
\left. \begin{array}{|c|c|c|c|}
\hline 
\multicolumn{2}{|c|}{} & \multicolumn{2}{|c|}{}\\
\multicolumn{2}{|c|}{\text{The foliated $BF$ theory}} 
&
\multicolumn{2}{|c|}{\text{The exotic $BF$ theory}} \\
\multicolumn{2}{|c|}{} & \multicolumn{2}{|c|}{} \\
\hline 
&&& \\
 \text{Gauge fields} & \text{Gauge} & \text{Gauge fields} & \text{Gauge} \\
 \text{and parameters} & \text{transformations} & \text{and parameters} & \text{transformations} \\
&&& \\
\hline 
&&&\\
B^i_0 - B^j_0  & \partial_0(\lambda^i - \lambda^j) & \hat{A}^{k(ij)}_0 & \partial_0 \hat{\alpha}^{k(ij)} \\
((i,j,k)=(1,2,3), & ((i,j,k)=(1,2,3), & ((i,j,k)=(1,2,3), & ((i,j,k)=(1,2,3),  \\
(2,3,1),(3,1,2)) & (2,3,1),(3,1,2)) &(2,3,1),(3,1,2)) & (2,3,1),(3,1,2))  \\
&&&\\
\hline 
&&&\\
B^i_k - B^j_k  & \partial_k (\lambda^i - \lambda^j) & \hat{A}^{ij} & \partial_k \hat{\alpha}^{k(ij)} \\
((i,j,k)=(1,2,3), & ((i,j,k)=(1,2,3), & ((i,j,k)=(1,2,3), & ((i,j,k)=(1,2,3),  \\
(2,3,1),(3,1,2)) & (2,3,1),(3,1,2)) &(2,3,1),(3,1,2)) & (2,3,1),(3,1,2))  \\
&&&\\
\hline 
&&&\\
\lambda^i - \lambda^j  & 2\pi m^i - 2\pi m^j & \hat{\alpha}^{k(ij)} & 2\pi \tilde{m}^i - 2\pi \tilde{m}^j \\
((i,j)=(1,2), & ((i,j)=(1,2), & ((i,j,k)=(1,2,3), & ((i,j)=(1,2),  \\
(2,3),(3,1)) & (2,3),(3,1)) &(2,3,1),(3,1,2)) & (2,3),(3,1))  \\
&&&\\
\hline 
&&&\\
m^k  &  & \tilde{m}^k &  \\
(k=1,2,3) & & (k=1,2,3) &  \\
&&&\\
\hline \end{array}\right. 
\end{align*}
\caption{The correspondences of the gauge fields and parameters between the foliated $BF$ theory and the exotic $BF$ theory in $3+1$ dimensions (the $B$-type fields).}
\label{3corresptable2}
\end{table}

%% file: BF-type_21.tex
\section{$BF$-type Theory in 2+1 Dimensions} \label{section 21}

In this section, we consider two $BF$-type theories in $2+1$ dimensions: %
a foliated $BF$ theory and an exotic $BF$ theory. Both of the theories are %
the continuum descriptions of the $\Z_N$ plaquette Ising model (see \cite{Johnston:2016mbz} for a review) and the $\Z_N$ lattice tensor gauge theory \cite{Seiberg:2020bhn}, both of which have subsystem symmetries and excitations of fractons. These two $BF$-type theories represent the same physics and we will show the explicit duality between them.

We take a three-torus of lengths $l^0$, $l^1$, $l^2$ as the spacetime %
and  the coordinates $(x^0,x^1,x^2)$ on it, with $x^0$ regarded as the Euclidean time. 
We consider theories that are Lorentz non-invariant and not fully rotation invariant. %
Instead, %
the spacetime symmetry is the spatial 90 degree rotational symmetry and the time translation %
as lattice models have.  
In the foliated theory the discrete rotational symmetry is not manifest, %
while in the exotic theory it is explicit.
In spite of the continuity of the spacetime, these theories can have discontinuous field configurations.

\subsection{2+1d Foliated $BF$ Theory} \label{2+1 fBF}

We will discuss a foliated $BF$ theory in $2+1$ dimensions. This is the $2+1$d version of the $3+1$d foliated QFT 
studied in \cite{Slagle:2018swq,Slagle:2020ugk,Hsin:2021mjn}. 

\subsubsection{Foliation and Foliated Gauge Fields} \label{ffield}

We consider a QFT on the $d$-dimensional manifold that is regarded as a stack of an infinite number of $(d-1)$-dimensional submanifolds. These submanifolds are called leaves and such a decomposition of a manifold is called a codimension-one foliation. A QFT on such a manifold is called a foliated QFT (FQFT) 
 \cite{Slagle:2020ugk}. 

A codimension-one foliation is characterized by a nonzero one-form foliation field $e$. The foliation field $e$ is orthogonal to the leaves of the foliation. For the foliation to be well-defined,  $e$ must satisfy the constraint %
\begin{align}
    e \wedge d e = 0\,.
\end{align}
The foliation field has a gauge redundancy under the transformation $e \rightarrow \gamma e$, where $\gamma$ is a scalar function. Using this redundancy, we can locally write the foliation field as $e = d f$, where $f$ is a scalar function. We can consider $f$ as a coordinate that specifies the leaves of the foliation. For example, we consider the flat foliation in $2+1$ dimensions that decomposes a $(x^1,x^2)$-plane into an infinite number of lines along the $x^1$ direction. Then the foliation field can be written as $e = d x^2$ locally. We can also consider multiple simultaneous foliations indexed by $k$ ($k=1,2,...,n_f$), where each foliation field is $e^k$. In the following, we consider the flat foliations $e^k = dx^k$.

A FQFT is a QFT coupled to foliation fields $e^k$ as backgrounds. A FQFT contains foliated gauge fields that can have discontinuous configurations. We %
consider two types of $U(1)$ foliated gauge fields for each foliation $k$ \cite{Hsin:2021mjn}. One is the foliated $A$-type $(1+1)$-form gauge field $\tilde{A}^k$ that obeys $\tilde{A}^k \wedge e^k = 0$.\footnote{The words $A$-type and $B$-type are the notation used only in this paper. In \cite{Slagle:2020ugk}, $A^k$ and $B^k$ denote what we call $\tilde{A}^k$ and $B^k$ in this paper, but in \cite{Hsin:2021mjn}, the symbols $A^k$ and $B^k$ are swapped compared to those in \cite{Slagle:2020ugk}.} $\tilde{A}^k$ can have one-form delta function singularities in the $x^k$ direction as $\delta(x^k - x^k_0)\, d x^k$. %
The gauge transformation of $\tilde{A}^k$ is
\begin{align}
    \tilde{A}^k \rightarrow \tilde{A}^k +d \tilde{\zeta}^k \, ,
\end{align}
where $\tilde{\zeta}^k$ is a $(0+1)$-form gauge parameter satisfying $\tilde{\zeta}^k \wedge e^k = 0$. 
The gauge parameter $\tilde{\zeta}^k$ has its own gauge transformation $\tilde{\zeta}^k \rightarrow \tilde{\zeta}^k + 2\pi d \xi^k$, where $\xi^k$ is a $x^k$-dependent function valued in integers. %
The gauge parameter $\tilde{\zeta}^k$ can have one-form delta function singularities in the $x^k$ direction, while the gauge parameters $\xi^k$ can have zero-form step function discontinuities $\theta(x^k -x^k_0)$ in the $x^k$ direction. For flat foliations $e^k = dx^k$, $\tilde{\zeta}^k$ can be locally written as $\zeta^k dx^k$, where $\zeta^k$ is a zero-form gauge parameter. The foliated $A$-type $(1+1)$-form gauge fields and its gauge parameters are summarized in Table~\ref{tablea}.
The other foliated gauge field is the foliated $B$-type gauge field $B^k$. In the foliated $BF$ theory in $2+1$ dimensions, $B^k$ is a zero-form gauge field that can have zero-form step function discontinuities in the $x^k$ direction. %
The gauge transformation of $B^k$ is 
\begin{align}
    B^k \rightarrow B^k +2\pi m^k - \mu \, ,
\end{align}
where the gauge parameter $m^k$ is a $x^k$-dependent function valued in integers and $\mu$ is a zero-form bulk gauge parameter. $m^k$ can have zero-form step function discontinuities in the $x^k$ direction. The foliated $B$-type zero-form gauge fields and its gauge parameters are summarized in Table~\ref{tableb}.

\begin{table}[htbp]
\centering
\begin{align*}
\left. \begin{array}{|c|c|c|c|}
\hline 
&&&\\
 \text{Gauge field} & \text{Constraints} & \text{Gauge transformations} & \text{Singularities and discontinuities} \\
 \text{and parameters} &&& \\
&&& \\
\hline 
&&&\\
(1+1)\text{-form}~ \tilde{A}^k & \tilde{A}^k \wedge e^k = 0 & \tilde{A}^k \rightarrow \tilde{A}^k +d \tilde{\zeta}^k & \text{one-form delta functions} \\
&&& \delta(x^k- x^k_0)\, d x^k \\
&&&\\
\hline 
&&&\\
(0+1)\text{-form}~ \tilde{\zeta}^k & \tilde{\zeta}^k \wedge e^k = 0 & \tilde{\zeta}^k \rightarrow \tilde{\zeta}^k + 2 \pi d \xi^k & \text{one-form delta functions} \\
&&& \delta(x^k- x^k_0)\, d x^k \\
&&&\\
\hline 
&&&\\
\text{$x^k$-dependent}  & && \text{zero-form step functions} \\
\text{function}~ \xi^k \in  \Z &&& \theta(x^k- x^k_0)  \\
&&&\\
\hline \end{array}\right. 
\end{align*}
\caption{The foliated $A$-type $(1+1)$-form gauge field and its gauge parameters.}
\label{tablea}
\end{table}

\begin{table}[htbp]
\centering
\begin{align*}
\left. \begin{array}{|c|c|c|}
\hline 
&&\\
 \text{Gauge field} & \text{Gauge transformations} & \text{Singularities and discontinuities} \\
 \text{and parameter} && \\
&& \\
\hline 
&& \\
\text{zero-form}~ B^k &   B^k \rightarrow B^k +2\pi m^k - \mu & \text{zero-form step functions} \\
&& \theta(x^k- x^k_0)  \\
&& \\ 
\hline 
&&\\
\text{$x^k$-dependent} && \text{zero-form step functions} \\
\text{function}~ m^k \in \Z && \theta(x^k- x^k_0)  \\
&&\\
\hline \end{array}\right. 
\end{align*}
\caption{The foliated $B$-type zero-form gauge field and its gauge parameters.}
\label{tableb}
\end{table}

\subsubsection{2+1d Foliated $BF$ Lagrangian} \label{21fbf}

The foliated $BF$ theory is a FQFT containing foliated gauge fields and bulk ordinary gauge fields %
with interactions among them.
The foliated $BF$ Lagrangian is
\begin{align}
\L_f = \sum^{n_f}_{k=1} \frac{iM_k}{2\pi} (dB^k + n_k b)\wedge \tilde{A}^k + \frac{iN}{2\pi} b \wedge da\,, \label{flagrangian}
\end{align}
where $\tilde{A}^k$ is an $A$-type $(1+1)$-form foliated gauge field satisfying $\tilde{A}^k \wedge e^k = 0$, $B^k$ is a $B$-type zero-form foliated gauge field, $a$ and $b$ are one-form gauge fields, and $N$, $M_k$, and $n_k$ are integers. These fields are $U(1)$ gauge fields and the gauge symmetry $U(1)$ is Higgsed down to $\Z_N$ or $\Z_{M_k}$. The first term $ \sum^{n_f}_{k=1} \frac{iM_k}{2\pi} dB^k\wedge \tilde{A}^k$ is a stack of $1+1$d $BF$ theories for each foliations, %
the third term $\frac{iN}{2\pi} b \wedge da$ is a bulk $2+1$d $BF$ theory, and the second term is interactions between the foliated fields and the bulk fields.

Let us discuss the special case where the foliations are flat, $n_f = 2$ (i.e., $e^k = dx^k$ for $k = 1,2$), $M_k=N$ and $n_k = 1$. In this case, the foliated gauge field $\tilde{A}^k$ can be written as $\tilde{A}^k = A^k \wedge dx^k$, where $A^k$ is a one-form gauge field. %
In this special case, the foliated Lagrangian can be written as
\begin{align}
    \L_f = \sum^{2}_{k=1} \frac{iN}{2\pi} (dB^k + b)\wedge A^k \wedge dx^k + \frac{iN}{2\pi} b \wedge da\,. \label{21folilag}
\end{align}

The equations of motion are
\begin{subequations}
\begin{align}
     \frac{N}{2\pi}(dB^k + b) \wedge dx^k = 0\,, \label{feom1}\\
     \frac{N}{2\pi}db = 0\,, \label{feom2}\\
     \frac{N}{2\pi}dA^k \wedge dx^k = 0\,, \label{feom3}\\
     \sum^{2}_{k=1} \frac{N}{2\pi} A^k \wedge dx^k + \frac{N}{2\pi} da =0\,. \label{feom4}
\end{align}
\end{subequations}

The gauge transformations are
 \begin{subequations}
 \begin{align}
     A^k \wedge dx^k &\rightarrow A^k \wedge dx^k + d\zeta^k \wedge dx^k\,, \label{fgauge1} \\
     B^k &\rightarrow B^k + 2\pi m^k - \mu\,,  \label{fgauge2} \\
     a &\rightarrow a + d\lambda - \sum^{2}_{k=1} \zeta^k dx^k\,, \label{fgauge3} \\
     b &\rightarrow b + d\mu\,, \label{fgauge4}
 \end{align}
 \end{subequations}
where $\zeta^k$, $m^k$ and $\mu$ are the gauge parameters explained in Section \ref{ffield}, and $\lambda$ are zero-form bulk gauge parameters that also have their own gauge transformations. 
The gauge transformation of $\lambda$ is $\lambda \rightarrow \lambda + 2\pi \xi^1 + 2\pi \xi^2$, where $\xi^k$ are $x^k$-dependent functions valued in integers explained in Section \ref{ffield}. Note that while $\xi^k$ are the parameters for the transformation of $\zeta^k$, the constant modes of $\xi^k$ do not affect $\zeta^k$ and rather make $\lambda$ a $U(1)$-valued function. The equations of motion and the gauge transformations imply that the bulk fields $a$, $b$ and their gauge parameters can have singularities and discontinuities. The singularities and discontinuities of $a$ are shown in Table~\ref{bulktable}, where $f^k_i$ and $g^k$ are some continuous functions with appropriate periodicity conditions.

\begin{table}[htbp]
\centering
\begin{align*}
\left. \begin{array}{|c|c|c|}
\hline 
&&\\
 \text{Gauge field} & \text{Gauge transformation} & \text{Terms including} \\
 \text{and parameter} && \text{singularities and discontinuities} \\
&& \\
\hline 
&& \\
 a_0 &   a_0 \rightarrow a_0 +\partial_0 \lambda  & f_0^1(x^0,x^2)\theta(x^1-x^1_0) +  f_0^2(x^0,x^1)\theta(x^2-x^2_0)  \\
&& \\ 
\hline 
&& \\
 a_1 &   a_1 \rightarrow a_1 +\partial_1 \lambda -\zeta^1  &  f^1_1(x^0,x^2)\delta(x^1-x^1_0) +  f^2_1(x^0,x^1)\theta(x^2-x^2_0)  \\
&& \\
\hline 
&& \\
 a_2 &   a_2 \rightarrow a_2 +\partial_2 \lambda -\zeta^2  &  f^1_2 (x^0,x^2)\theta(x^1-x^1_0) + f^2_2 (x^0,x^1)\delta(x^2-x^2_0)    \\
&& \\
\hline 
&& \\
 \lambda &   \lambda \rightarrow \lambda + 2\pi \xi^1 + 2\pi \xi^2  &  g^1(x^0,x^2)\theta(x^1-x^1_0) + g^2(x^0,x^1)\theta(x^2-x^2_0)  \\
&& \\
\hline \end{array}\right. 
\end{align*}
\caption{Singularities and discontinuities of the bulk gauge field $a$ and its gauge parameter.}
\label{bulktable}
\end{table}

Integrating the fields out and considering specific field configurations, we can show that the following quantities are quantized: %
\begin{subequations}
    \begin{align}
    \oint_{C_1^0} a \in \frac{2\pi}{N}\Z\,, \label{aqua} \\ 
    \oint_{S_2^k} A^k \wedge dx^k \in  \frac{2\pi}{N}\Z\,, \label{akqua} \\
    \oint_{C_1} b \in  \frac{2\pi}{N}\Z\,, \label{bqua} \\
    B^1-B^2 \in \frac{2\pi}{N}\Z\,, \label{bkqua}
\end{align}
\end{subequations}
where $C_1^0$ is a closed one-dimensional loop along the time $x^0$ direction, %
$C_1$ is %
an arbitrary closed one-dimensional loop, and $S_2^k$ is a two-dimensional strip with a fixed width %
along the $x^k$ direction. %
For example for \eqref{akqua}, %
there is a configuration %
\begin{align}
    B^1 = 2\pi j \frac{x^2}{l^2} (\theta(x^1-x^1_1) - \theta(x^1-x^1_2) ) \,,
\end{align}
where $j$ is an integer. This configuration is periodic in $x^2$ up to the gauge transformation \eqref{fgauge2}. With this configuration, we have
\begin{equation}
\begin{split}
    & \oint_{C^0_1 \times C^1_1 \times C^2_1}   d B^1  \wedge A^1 \wedge dx^1  \\
    &= \oint_{C^0_1 \times C^1_1 \times C^2_1} 2\pi j \left(\theta(x^1-x^1_1) - \theta(x^1-x^1_2) \right) \frac{1}{l^2}  dx^2  \wedge A^1 \wedge dx^1 \\
     &= \frac{2\pi j}{l^2} \oint_{C^2_1} dx^2 \oint_{C_1^0 \times [ x^1_1 , x^1_2]} A^1 \wedge dx^1  \,,
\end{split}
\end{equation}
where  $C_1^2$ is a closed one-dimensional loop along the $x^2$ direction. %
If we use the equation of motion (\ref{feom3}) and perform the sum over $j$ in this configuration as a part of the path-integral in terms of $B^1$, we get
\begin{align}
    \sum_{j \in \Z}\exp \left[ \frac{i N j}{l^2} \oint_{C^2_1} dx^2 \oint_{C_1^0 \times [ x^1_1 , x^1_2]} A^1 \wedge dx^1  \right] =  \sum_{j \in \Z}\exp \left[ i N j \oint_{C_1^0 \times [ x^1_1 , x^1_2]} A^1 \wedge dx^1  \right]\,.
\end{align}
Then $N \oint_{C_1^0 \times [ x^1_1 , x^1_2]} A^1 \wedge dx^1$ must be in $2\pi \Z$; the configuration of $A^1$ not satisfying this condition does not contribute to the path integral. %
Again from the equation of motion (\ref{feom3}), $C_1^0$ can be deformed into $C_1^{02}$ that is a closed loop in the $(x^0,x^2)$-plane.

\subsubsection{Gauge-Invariant Operators}

Let us consider %
the gauge-invariant operators, %
which describe excitations moving in spacetime.

The first one is
\begin{align}
    F^q[C_1^0] = \exp \left[ i q\oint_{C_1^0} a \right]\,, \label{ffracton}
\end{align}
where $q$ is an integer. From (\ref{aqua}), we can see that $F^N[C_1^0] = 1$, and thus $F^q[C_1^0]$ is a $\Z_N$ operator: $F^{q+N} =  F^q$.
The deformation of $C_1^0$ would break the gauge invariance of the operator under the transformation $\zeta^k$. If the contour were $C_1^{02}$ in the $(x^0,x^2)$-plane, under the gauge transformation of $a$, the defect operator would become 
\begin{align}
    F^q[C^{02}_1] \rightarrow  \exp \left[ i q \oint_{C^{02}_1} \left\{ dx^0 \partial_0 \lambda  + dx^2 ( \partial_2 \lambda - \zeta^2) \right\} \right] F^q[C^{02}_1]\,,
\end{align}
which would not be gauge invariant.
Since $C_1^0$ is a line in the time direction, this one-dimensional operator is the defect operator that describes a fracton, which cannot move in space.

The second one is
\begin{align}
    V^q[x] = \exp \left[ i q(B^1-B^2) \right] \label{fpoint}\,,
\end{align}
where $q$ is an integer again. From (\ref{bkqua}), we can see that $V^N[x] = 1$ and thus $V^q[x]$ is also a $\Z_N$ operator: $V^{q+N} = V^q$. The point operator $V^q[x]$ is the symmetry operator that generates a $\Z_N$ electric global symmetry, which is a subsystem symmetry.

The third ones are 
\begin{align}
    W_k^q[S^k_2] = \exp \left[ i q \oint_{S_2^k} A^k \wedge dx^k \right] \,, \quad k=1,2\,,  \label{fstrip}
\end{align}
where $q$ is an integer again. Similarly from (\ref{akqua}), $W_k^q[S^k_2]^N = 1$ and thus $W_k^q[S^k_2]$ are
$\Z_N$ operators: $W^{q+N} = W^q$. These two-dimensional strip operators describe a dipole of fractons separated in the $x^k$ direction, which can move in the other direction in space, like a lineon. If $S^k_2$ are in the $(x^1,x^2)$-plane, these operators become the symmetry operators that generate $\Z_N$ dipole global symmetries, which are also subsystem symmetries. %

These two types of symmetry operators are the charged objects under the other symmetry. That is,
$V^p[x]$ and $W_k^q[S^k_2]$ satisfy the following relations at equal time:
\begin{align}
    V^p[x] \ W_k^q[S^k_2] =  \mathrm{e}^{2\pi i p q /N} \  W_k^q[S^k_2]  \ V^p[x]\,, \quad \text{if} \quad x^k_1 < x^k < x^k_2\,,
\end{align}
when $S^k_2$ is $[x^1_1, x^1_2] \times C_1^2$ $(k = 1)$ or $C_1^1 \times [x^2_1, x^2_2]$ $(k = 2)$, where $C_1^r\, (r=1,2)$ is a closed loop in the $x^r$-plane. 
We can derive this relation using the canonical commutation relation
\begin{subequations}
\begin{align}
    \left[ B^1(x^0, x^1, x^2) , A^1_2 (x^0, y^1, y^2) \right] &= -\frac{2\pi i}{N} \delta^2(x^1 - y^1, x^2 - y^2) \,, \\
    \left[ B^2(x^0, x^1, x^2) , A^2_1 (x^0, y^1, y^2) \right] &= + \frac{2\pi i}{N} \delta^2(x^1 - y^1, x^2 - y^2) \,.
\end{align}
\end{subequations}
All the other commutators are zero.

In addition, the bulk $2+1$d $BF$ theory has a gauge-invariant operator
\begin{align}
    T^q[C_1] = \exp \left[ iq \oint_{C_1} b\right]\,. \label{bop}
\end{align}
From (\ref{bqua}), this $b$ operator is also a $\Z_N$ operator: $T^{q+N} = T^q$. %
This operator has the winding action on the gauge-invariant operator (\ref{ffracton}) as
\begin{align}
 T^p[C_1]\  \cdot F^q[C_1^0] = \mathrm{e}^{-2\pi i p q /N} F^q[C_1^0]\,,
\end{align}
when $C_1$ surrounds $C_1^0$ \cite{Maldacena:2001ss,Banks:2010zn,Kapustin:2014gua,Gukov:2013zka}. 
Without the defect operator $F^q$ inside $C_1$, the $b$ operator $T^q$ becomes trivial, which corresponds to a time-like symmetry \cite{Gorantla:2022eem}.\footnote{A time-like symmetry acts nontrivially on a Hilbert space in the presence of time-like defects. Without the defect operator, the time-like symmetry operator becomes trivial. We thank Pranay Gorantla for his comments on the relations between the $b$ operators and the time-like symmetries.} 
For the later purpose, it will be convenient to consider the case when $C_1$ is a rectangle $C_1^{12,\text{rect}}(x^1_1,x^1_2,x^2_1,x^2_2)$ in the space. In this case,
using the equation of motion \eqref{feom1}, 
the integral in the definition of $T^q$ can be performed as
\begin{equation}
\begin{split}
        T^q\left[C_1^{12,\text{rect}}(x^1_1,x^1_2,x^2_1,x^2_2)\right] &= \exp\left[ 
i q \oint_{C_1^{12,\text{rect}}} (-\partial_1 B^2 dx^1 - \partial_2 B^1 dx^2 ) \right] \\
        &= \exp\left[-i q \Delta_{12} (B^1 - B^2)(x^1_1,x^1_2,x^2_1,x^2_2)    \right] \,, \label{timelikebop}
\end{split}
\end{equation}
where $\Delta_{12} f(x^1_1,x^1_2,x^2_1,x^2_2) = f(x^1_2,x^2_2) - f(x^1_2,x^2_1) - f(x^1_1,x^2_2) + f(x^1_1,x^2_1)$. This quadrupole operator is a product of the gauge-invariant operators $V^q$ localized at the corners of the rectangle.

Note that the operator $T^q$ cannot be remotely detected by an operator within a spatial slice, as the fracton operator $F^q$ cannot be bent to braid with $T^p$. %
This implies that there is no physical excitation corresponding to the operator $T^p$ in this situation.
This contrasts with the case of usual topological field theory where every non-trivial line operator corresponds to a physical excitation.

\subsection{2+1d Exotic $BF$ Theory} \label{2+1 eBF}

In this section, we review the exotic $BF$ theory in $2+1$ dimensions, which is the $\Z_N$ tensor gauge theory in \cite{Seiberg:2020bhn}. %
In Section \ref{2+1 corr} we will see that this exotic theory is equivalent to the foliated $BF$ theory discussed in the previous section.

\subsubsection{Tensor Gauge Fields} \label{tensor}
We will discuss an exotic theory that is not Lorentz invariant and has only the $90$ degree rotational invariance. %
Such theories can have tensor gauge fields, %
each of which is in a representation of the $90$ degree rotation group $\Z_4$. Irreducible representations of $\mathbb{Z}_4$ are one-dimensional ones $\bm{1}_n\, (n=0, \pm 1, 2)$, where $n$ is the spin. 
The exotic $BF$ theory contains %
a compact scalar $\phi^{12}$ in the representation $\bm{1}_2$ and a $U(1)$ tensor gauge field $(A_0,A_{12})$ in the representation $(\bm{1}_0,\bm{1}_2)$. Their gauge transformations are
\begin{subequations}
\begin{align}
    A_{0} &\rightarrow A_{0} + \partial_0 \alpha \,, \label{egauge1} \\
    A_{12} &\rightarrow A_{12} + \partial_1\partial_2 \alpha \,, \label{egauge2} \\
    \phi^{12} &\rightarrow \phi^{12} + 2\pi \tilde{m}^1 - 2\pi \tilde{m}^2 \,, \label{egauge3}
\end{align}
\end{subequations}
where $\alpha$ is a $U(1)$-valued gauge parameter: $\alpha \sim \alpha + 2\pi$, in the representation $\bm{1}_0$, and $\tilde{m}^k$ are $x^k$-dependent functions valued in integers. The gauge parameter $\alpha$ has its own gauge transformation: $\alpha \rightarrow \alpha + 2\pi \tilde{n}^1 + 2\pi \tilde{n}^2$, where $\tilde{n}^k$ are $x^k$-dependent functions valued in integers. 
$A_{12}$ can have delta function singularities and $A_0$, $\alpha$ and $\phi^{12}$ can have step function discontinuities as in Table~\ref{exotictable}, where $\tilde{f}_0^k$, $\tilde{f}_{12}^k$, $\tilde{g}^k$,  and $\tilde{h}^k$ are some continuous functions  with appropriate periodicity conditions. 
For example, the following configurations are allowed:
\begin{align}
    A_{12} = 2\pi \frac{x^0}{l^0} \left[ \frac{1}{l^2} \delta(x^1-x^1_0) + \frac{1}{l^1} \delta(x^2-x^2_0) - \frac{1}{l^1 l^2} \right] \,,
\end{align}
\begin{align}
    \phi^{12} = 2\pi  \left[ \frac{x^2}{l^2} \theta(x^1-x^1_0) + \frac{x^1}{l^1} \theta(x^2-x^2_0) - \frac{x^1 x^2}{l^1 l^2} \right] \,.
\end{align}
The gauge parameters $\tilde{m}^k$ and $\tilde{n}^k$ can have step function discontinuities in the $x^k$ direction.

\begin{table}[htbp]
\centering
\begin{align*}
\left. \begin{array}{|c|c|c|}
\hline 
&&\\
 \text{Gauge fields} & \text{Gauge transformation} & \text{Terms including} \\
 \text{and parameter} && \text{singularities and discontinuities} \\
&& \\
\hline 
&& \\
 A_0 &   A_0 \rightarrow A_0 +\partial_0 \alpha  & \tilde{f}_0^1(x^0,x^2)\theta(x^1-x^1_0) +  \tilde{f}_0^2(x^0,x^1)\theta(x^2-x^2_0)  \\
&& \\ 
\hline 
&& \\
 A_{12} &   A_{12} \rightarrow A_{12} +\partial_1 \partial_2 \alpha   &  \tilde{f}_{12}^1(x^0,x^2)\delta(x^1-x^1_0) +  \tilde{f}_{12}^2(x^0,x^1)\delta(x^2-x^2_0)  \\
&& \\
\hline 
&& \\
 \alpha &  \alpha \rightarrow \alpha + 2\pi \tilde{n}^1 + 2\pi \tilde{n}^2  & \tilde{g}^1(x^0,x^2)\theta(x^1-x^1_0) +  \tilde{g}^2(x^0,x^1)\theta(x^2-x^2_0)  \\
&& \\ 
\hline
&& \\
 \phi^{12} &  \phi^{12} \rightarrow \phi^{12} + 2\pi \tilde{m}^1 - 2\pi \tilde{m}^2  &  \tilde{h}^1 (x^0,x^2)\theta(x^1-x^1_0) + \tilde{h}^2(x^0,x^1)\theta(x^2-x^2_0)    \\
&& \\
\hline \end{array}\right. 
\end{align*}
\caption{Singularities and discontinuities of the tensor gauge fields and their gauge parameters.}
\label{exotictable}
\end{table}

\subsubsection{2+1d Exotic $BF$ Lagrangian}

The exotic $BF$ Lagrangian is
\begin{align}
    \L_e = \frac{i N}{2\pi}\phi^{12}(\partial_0 A_{12} - \partial_{1}\partial_{2} A_0) \,. \label{elagrangian}
\end{align}

The equations of motion are
\begin{subequations}
\begin{align}
    \frac{N}{2\pi} \partial_{1}\partial_{2} \phi^{12} = 0 \,, \label{eeom1}\\
    \frac{N}{2\pi} \partial_{0} \phi^{12} = 0 \,, \label{eeom2} \\
    \frac{N}{2\pi} (\partial_0 A_{12} - \partial_{1}\partial_{2} A_0) = 0 \,. \label{eeom3}
\end{align}
\end{subequations}

Integrating specific configurations %
 out, we can show that the following quantities are quantized:
\begin{subequations}
\begin{gather}
    \oint_{C_1^0} dx^0 A_0 \in \frac{2\pi}{N}\Z \,, \label{eaqua} \\ 
    \oint_{S^1_2} ( dx^0 dx^1 \partial_1 A_0 + dx^2 dx^1 A_{12} )  \in  \frac{2\pi}{N}\Z \,, \label{eakqua1} \\
    \oint_{S^2_2} ( dx^0 dx^2 \partial_2 A_0 + dx^1 dx^2 A_{12} ) \in  \frac{2\pi}{N}\Z \,, \label{eakqua2} \\
    \phi^{12} \in \frac{2\pi}{N}\Z\,, \label{phiqua}
\end{gather}
\end{subequations}
where $C_1^0$ is a closed one-dimensional loop along the time $x^0$ direction, and $S^k_2$ is a two-dimensional strip with a fixed width extended along the $x^k$ direction. For example for \eqref{eakqua1}, there is a configuration
\begin{align}
    \phi^{12} = 2\pi j \frac{x^2}{l^2} \left[ \theta (x^1-x^1_1) - \theta (x^1-x^1_2) \right] \,,
\end{align}
where $j$ is an integer. Then we have
\begin{align}
    \oint_{C^1_1} dx^1 \oint_{C^2_1} dx^2 \partial_2 \phi^{12} \partial_1 A_0 = 2\pi j \oint_{C^1_1}  dx^1 ( \theta(x^1 - x^1_1) - \theta(x^1 - x^1_2) ) 
 \frac{1}{l^2}  \oint_{C^2_1} dx^2   \partial_1 A_0 \,,
\end{align}
 Integrating this configuration out and using the equation of motion (\ref{eeom3}), this part of partition function becomes
\begin{align}
     \sum_{j \in \Z} \exp \left[ \frac{i N j}{l^2}  \oint_{C^2_1} dx^2 \oint_{C^0_1 \times [x^1_1, x^1_2]} dx^0 dx^1 \partial_1 A_0 \right] = \sum_{j \in \Z} \exp \left[ i N j  \oint_{C^0_1 \times [x^1_1, x^1_2]} dx^0 dx^1 \partial_1 A_0 \right]\,,
\end{align}
and the term $N \oint_{C^0_1 \times [x^1_1, x^1_2]} dx^0 dx^1 \partial_1 A_0 $ must be an integer %
for the configuration to contribute. Again using the equation of motion (\ref{eeom3}), $C^0_1$ can be deformed into $C^{02}_1$ and the term becomes $\oint_{C^{02}_1 \times [x^1_1, x^1_2]} ( dx^0 dx^1 \partial_1 A_0 + dx^2 dx^1  A_{12} )$.

\subsubsection{Gauge-Invariant Operators}

Let us discuss gauge-invariant operators. The defect operator that describes fractons is
\begin{align}
    \tilde{F}^q[C_1^0] = \exp \left[ i q \oint_{C_1^0} dx^0 A_0\right] \,. \label{efracton}
\end{align}
As in the case of the foliated $BF$ theory, the deformation of $C^0_1$ would break the gauge invariance of the operator.
The symmetry operator that generates a $\Z_N$ electric global symmetry is
\begin{align}
    \tilde{V}^q[x] = \exp \left[ i q\phi^{12} \right]\,. \label{epoint}
\end{align}
The strip operators that describe a dipole of fractons are
\begin{subequations}
\begin{align}
    \tilde{W}_1^q\left[ S^1_2 \right] = \exp \left[ i q \oint_{S^1_2} ( dx^0 dx^1 \partial_1 A_0 + dx^2 dx^1 A_{12} )  \right] \,, \label{estrip1}\\
    \tilde{W}_2^q\left[ S^2_2  \right] = \exp \left[ i q \oint_{S^2_2} ( dx^0 dx^2 \partial_2 A_0 + dx^1 dx^2 A_{12} )  \right]\,. \label{estrip2}
\end{align}
\end{subequations}
If $S^k_2$ are in the $(x^1,x^2)$-plane, $\tilde{W}_k^q$ are the symmetry operators that generate $\Z_N$ dipole global symmetries. 
These gauge-invariant operators are $\Z_N$ operators: $q$ is an element of $\Z_N$. 

The two types of symmetry operators satisfy the following relations
\begin{align}
    \tilde{V}^p[x] \ \tilde{W}_k^q[S^k_2] =  \mathrm{e}^{2\pi i p q  /N} \  \tilde{W}_k^q[S^k_2]  \ \tilde{V}^p[x]\ ,\quad \text{if} \quad x^k_1 < x^k < x^k_2\,.
\end{align}
We can derive this relation using the canonical commutation relations at equal time:
\begin{align}
    \left[ \phi^{12}(x^0, x^1, x^2) , A_{12} (x^0, y^1, y^2) \right] &= -\frac{2\pi i}{N} \delta^2(x^1 - y^1, x^2 - y^2) \,.
\end{align}
All the other commutators are zero. 
These symmetries in the exotic $BF$ theory have the same structure as the foliated $BF$ theory discussed in Section \ref{2+1 fBF}.

In addition, there is a gauge-invariant operator that can detect the fracton defect operator:
\begin{align}
    \tilde{T}^q\left[C_1^{12,\text{rect}}(x^1_1,x^1_2,x^2_1,x^2_2)\right] = \exp\left[-i q \Delta_{12} \phi^{12} (x^1_1,x^1_2,x^2_1,x^2_2) \label{etimelikeop}   \right] \,.
\end{align}
This quadrupole operator is a product of the gauge-invariant operators $\tilde{V}^q$ localized at the corners of the rectangle, which is a time-like symmetry \cite{Gorantla:2022eem}. The operator $\tilde{T}^p$ can detect the fracton operator $\tilde{F}^q$:
\begin{align}
    \tilde{T}^p\left[C_1^{12,\text{rect}}(x^1_1,x^1_2,x^2_1,x^2_2)\right] \cdot \tilde{F}^q[C^0_1] = \mathrm{e}^{-2\pi i p q/N} \tilde{F}^q[C^0_1] \,,
\end{align}
when $C_1^{12,\text{rect}}(x^1_1,x^1_2,x^2_1,x^2_2)$ surrounds $C^0_1$.

\subsection{Correspondences in 2+1 Dimensions} \label{2+1 corr}

The $2+1$d foliated $BF$ theory explained in Section \ref{2+1 fBF} and the $2+1$d exotic $BF$ theory explained in Section \ref{2+1 eBF} are equivalent in case that $e^k = dx^k$, $M_k=N$ and $n_k = 1$ $(k = 1,2)$, which we call the foliated-exotic duality. We identify the gauge-invariant operators in the foliated $BF$ theory with those of the exotic $BF$ theory. %
By matching the gauge-invariant operators, we can derive the correspondences of the gauge fields and parameters. 

As noted in \cite{Hsin:2021mjn}, some foliated theories with the flat condition 
\begin{align}
    d \left( \sum^{2}_{k=1} \frac{N}{2\pi} A^k \wedge dx^k \right) = 0 \,. \label{flatcond}
\end{align}
correspond to the tensor gauge theories. In the foliated $BF$ theory explained in Section \ref{2+1 fBF}, integrating $b$ out leads to the equation of motion (\ref{feom4}) that becomes the flat condition \eqref{flatcond}. %
In the following, we will explicitly see how the gauge-invariant operators are identified and the gauge fields and parameters match under the condition \eqref{flatcond}. The precise correspondences include the bulk gauge field $a$ in the FQFT side in a non-trivial way. We also see that the allowed singularities and discontinuities of the fields and parameters match between the two sides.

First, let us consider the fracton defect operators. %
We identify the operators $F^q[C_1^0]$ with $\tilde{F}^q[C_1^0]$ defined in \eqref{ffracton} and \eqref{efracton}:
\begin{align}
    \exp \left[ i q\oint_{C_1^0} a \right] \simeq \exp \left[ i q \oint_{C_1^0} dx^0 A_0\right] \,,
\end{align}
which leads to the field correspondence
\begin{align}
    a_0 \simeq A_0 \,.
    \label{a0A0}
\end{align}
The gauge transformations of $a_0$ and $A_0$ explained in (\ref{fgauge3}) and (\ref{egauge1}) are
\begin{subequations}
    \begin{align}
        a_0 &\rightarrow a_0 + \partial_0 \lambda \,, \\
        A_0 &\rightarrow A_0 + \partial_0 \alpha \,,
    \end{align}
\end{subequations}
from which we obtain the gauge parameter correspondence
\begin{align}
\lambda \simeq \alpha \,. \label{gcorr1}
\end{align}
Moreover, the gauge transformations of $\lambda$ and $\alpha$ are 
\begin{subequations}
    \begin{align}
        \lambda &\rightarrow \lambda+ 2\pi \xi^1 + 2\pi \xi^2 \,, \\
        \alpha &\rightarrow \alpha + 2\pi \tilde{n}^1 + 2\pi \tilde{n}^2\,,
    \end{align}
\end{subequations}
which can be matched by 
\begin{align}
         \xi^k \simeq  \tilde{n}^k  \,.
\end{align}
In these correspondences, one can check that their singularities and discontinuities are also matched.

The equations of motion (\ref{feom4}) in components are 
\begin{subequations}
\begin{align}
 \frac{N}{2\pi} ( A^1_0 + \partial_0 a_1 - \partial_1 a_0 ) &= 0 \,, \\
 \frac{N}{2\pi} ( A^2_0 + \partial_0 a_2 - \partial_2 a_0 ) &= 0 \,, \\
 \frac{N}{2\pi} ( A^1_2 - A^2_1  + \partial_2 a_1 - \partial_1 a_2 ) &= 0 \,. \label{beomc3}
\end{align}
\end{subequations}
These equations of motion, when combined with \eqref{a0A0}, imply
\begin{subequations}
\begin{align}
    A^1_0 + \partial_0 a_1 &\simeq \partial_1 A_0 \,, \label{corr2}\\
    A^2_0 + \partial_0 a_2 &\simeq \partial_2 A_0\,. \label{corr3}
\end{align}
\end{subequations}
Note that the gauge transformations by $\zeta^k$ in the left hand sides cancel out, and thus these are consistent with  the correspondence \eqref{gcorr1}.

Next, let us consider the strip operators. We want to identify the operators $W^q_k[S^k_2]$ with $\tilde{W}^q_k[S^k_2]$ defined in (\ref{fstrip}), (\ref{estrip1}) and (\ref{estrip2}), but the gauge transformations of the exponents are not matched and the field correspondences would be inconsistent with (\ref{corr2}) and (\ref{corr3}). Therefore, we define the modified gauge-invariant strip operators $W^q_{k,\text{mod}}[S^k_2]$ as
\begin{align}
    W^q_{k,\text{mod}}[S^k_2] = \exp \left[ i q \oint_{S^k_2} \left( A^k  \wedge dx^k + d(a_k dx^k) \right) \right] \,, \quad k=1,2 \,, \label{modstrip}
\end{align}
where the exponents are quantized:
\begin{align}
    \oint_{S^k_2} \left( A^k  \wedge dx^k + d(a_k dx^k) \right) \in \frac{2\pi}{N} \Z \,, \label{quamod}
\end{align}
and therefore $W^q_{k,\text{mod}}[S^k_2]$ are $\Z_N$ operators. For example for $k=1$, to show this quantization, we consider a configuration
\begin{align}
    b = 2\pi j \frac{1}{l^2} ( \theta(x^1-x^1_1) - \theta(x^1-x^1_2) ) dx^2 \,,
\end{align}
where $j$ is an integer. Using the equation of motion \eqref{feom3}, we have
\begin{equation}
\begin{split}
    & \oint_{C^0_1 \times C^1_1 \times C^2_1}   b  \wedge \left( \sum_{k=1}^2 A^k \wedge dx^k + da \right)  \\
    &= \frac{2\pi j}{l^2} \oint dx^2 \oint_{C_1^0 \times [ x^1_1 , x^1_2]} ( A_0^1 + \partial_0 a_1 - \partial_1 a_0 )\, dx^0 \wedge dx^1 \\
    &= 2\pi j  \oint_{C_1^0 \times [ x^1_1 , x^1_2]} ( A_0^1 + \partial_0 a_1 - \partial_1 a_0 )\, dx^0 \wedge dx^1\,.
\end{split}
\end{equation}
From the quantization \eqref{aqua}, the term $ \oint_{C^0_1} a_0 dx^0$ is in $2\pi\Z/N$. Integrating this configuration out, this part of partition function becomes
\begin{equation}
\begin{split}
    & \sum_{j \in \Z} \exp \left[ i N j \oint_{C_1^0 \times [ x^1_1 , x^1_2]} ( A_0^1 + \partial_0 a_1 )\, dx^0 \wedge dx^1 \right] \\
    &= \sum_{j \in \Z} \exp \left[ i N j \oint_{C_1^0 \times [ x^1_1 , x^1_2]} ( A^1 \wedge dx^1 + d (a_1 dx^1) ) \, \right] \,.
\end{split}
\end{equation}
Then $N \oint_{C_1^0 \times [ x^1_1 , x^1_2]} ( A^1 \wedge dx^1 + d (a_1 dx^1) )$ must be in $2\pi \Z$. Again using the equation of motion \eqref{feom3}, we can deform the $C^0_1$ into $C^{02}$ and conclude the quantization \eqref{quamod}.

Having prepared the operator \eqref{modstrip}, we identify the operators $W'^q_k[S^k_2]$ with $\tilde{W}^q_k[S^k_2]$:
\begin{subequations}
\begin{align}
    \exp \left[ i q \oint_{S^1_2} \left( A^1  \wedge dx^1 + d(a_1 dx^1) \right) \right]  &\simeq 
    \exp \left[ i q \oint_{S^1_2} ( dx^0 dx^1 \partial_1 A_0 + dx^2 dx^1 A_{12} )  \right] \,, \\
    \exp \left[ i q \oint_{S^2_2} \left( A^2  \wedge dx^2 + d(a_2 dx^2) \right) \right]  &\simeq 
    \exp \left[ i q \oint_{S^2_2} ( dx^0 dx^2 \partial_2 A_0 + dx^1 dx^2 A_{12} )  \right] \,,
\end{align}
\end{subequations}
which leads to the field correspondences
\begin{subequations}
\begin{align} 
    A^1_2 + \partial_2 a_1 &\simeq A_{12} \,, \\
    A^2_1 + \partial_1 a_2 &\simeq A_{12} \,,
\end{align}
\end{subequations}
and also %
(\ref{corr2}) and (\ref{corr3}) again. %
The terms $\partial_2 a_1$ and $\partial_1 a_2$ make the gauge transformations match with those of $A_{12}$ under  the gauge parameter correspondence (\ref{gcorr1}). 

Note that the ratio of $W_{k,\text{mod}}^q$ to $W_k^q$,
\begin{equation}
    W^q_{k,\text{mod}}[S^k_2](W^q_{k}[S^k_2])^{-1} = \exp \left[ i q \oint_{S^k_2}  d(a_k dx^k) \right] \,,
        \label{modstripdiff}
\end{equation}
is trivial when $a_k$ is single-valued on $S^k_2$ by applying the Stokes' theorem. This means that there is no operator that braids with the ratio operator and thus it does not describe a physical excitation.
This subtlety is tied to the fractonic nature of the system since in a non-fractonic topological field theory every operator corresponds to a physical excitation\cite{Kong:2014qka,Johnson-Freyd:2020usu}.%

Lastly, let us consider the $\Z_N$ electric global symmetry operators. We identify $V^q[x]$ with $\tilde{V}^q[x]$ defined in (\ref{fpoint}) and (\ref{epoint}):
\begin{align}
    \exp \left[ i q ( B^1 - B^2 ) \right] \simeq \exp \left[ i q \phi^{12} \right] \,.
\end{align}
Then we can derive the field correspondence
\begin{align}
    B^1 - B^2 \simeq \phi^{12} \,. \label{corrbphi}
\end{align}
The gauge transformations by $\mu$ cancel out in the left-hand side. From the gauge transformations (\ref{fgauge2}) and (\ref{egauge3}), we obtain
\begin{align}
    m^1 -m^2 \simeq \tilde{m}^1 - \tilde{m}^2 \,.
\end{align}
Considering the discontinuities, we can see that
\begin{align}
    m^1 \simeq \tilde{m}^1\,, \quad m^2 \simeq \tilde{m}^2 \,.
\end{align}
Under the correspondence \eqref{corrbphi}, the time-like symmetry operator $T^q[C_1^{12,\text{rect}}]$ defined \eqref{timelikebop} corresponds to $\tilde{T}^q[C_1^{12,\text{rect}}]$ defined in  \eqref{etimelikeop}. Note that on a Hilbert space with fracton defect operators, the $b$ operator $T^q[C_1]$ is a product of $T^q[C_1^{12,\text{rect}}]$ surrounding the defects that are surrounded by $C_1$.

Under these correspondences, the equations of motion are also matched. %
Moreover, after integrating $b$ out, and then using the correspondences, the Lagrangians (\ref{21folilag}) and (\ref{elagrangian}) are exactly matched.

%% file: BF-type_31.tex
\section{$BF$-type Theory in 3+1 Dimensions} \label{section31}

In this section, we  %
consider a foliated $BF$ theory and an exotic $BF$ theory in $3+1$ dimensions. Both of the theories are the continuum descriptions of the $\Z_N$ X-cube model \cite{Vijay:2016phm} and the $\Z_N$ lattice tensor gauge theory \cite{Seiberg:2020cxy}. 
As in the case of $2+1$ dimensions, these two $BF$-type theories represent the same physics and we will show the explicit duality between them. Basically the discussion proceeds in parallel with that in $2+1$ dimensions in Section \ref{section 21}.

We take a four-torus of lengths $l^0$, $l^1$, $l^2$, $l^3$ as a spacetime and the coordinates $(x^0,x^1,x^2,x^3)$ on it, with $x^0$ regarded as the Euclidean time. The spatial symmetry is the $S_4$ group generated by the 90 degree rotations along one of the axes, as the cubic lattice has. %

\subsection{3+1d Foliated $BF$ Theory} \label{31fbf}

We review the foliated $BF$ theory in $3+1$ dimensions\cite{Slagle:2018swq,Slagle:2020ugk,Hsin:2021mjn}.

\subsubsection{Foliated Gauge Fields} \label{3ffield}

In the foliated $BF$ theory in $3+1$ dimensions, the foliated $A$-type $(1+1)$-form gauge field $\tilde{A}^k$ is almost the same as $(2+1)$-dimensional one, while the foliated $B$-type gauge field $B^k$ is a one-form gauge field that can have zero-form step function discontinuities in the $x^k$ direction. The gauge transformation of $B^k$ is 
\begin{align}
    B^k \rightarrow B^k +d \lambda^k + \tilde{\beta}^k - \mu \, ,
\end{align}
where $\lambda^k$ is a zero-form gauge parameter, $\tilde{\beta}^k$ is a $(0+1)$-form gauge parameter satisfying $\tilde{\beta}^k \wedge e^k = 0$, and $\mu$ is a one-form bulk gauge parameter. The gauge parameter $\lambda^k$ has its own gauge transformation $\lambda^k \rightarrow \lambda^k + 2\pi m^k + \nu$, where $m^k$ is a $x^k$-dependent function valued in integers, $\beta^k$ also has its own gauge transformation $\tilde{\beta}^k \rightarrow \tilde{\beta}^k + 2\pi d m^k$, and $\mu$ also has its own gauge transformation $\mu \rightarrow \mu + d\nu$, where $\nu$ is a zero-form gauge parameter. %
The gauge parameters $\lambda^k$ and $m^k$ can have zero-form step function discontinuities in the $x^k$ direction. The gauge parameters $\tilde{\beta}^k$ can have one-form delta function singularities in the $x^k$ direction that cancel out the delta function terms in $d\lambda^k$. The foliated $B$-type one-form gauge fields and its gauge parameters are summarized in Table \ref{3tableb}.

\begin{table}[htbp]
\centering
\begin{align*}
\left. \begin{array}{|c|c|c|c|}
\hline 
&&&\\
 \text{Gauge field} & \text{Constraint} & \text{Gauge transformation} & \text{Singularities and Discontinuities} \\
 \text{and parameters} &&& \\
&&& \\
\hline 
&&&\\
\text{one-form}~ B^k &&   B^k \rightarrow B^k +d \lambda^k + \tilde{\beta}^k - \mu & \text{zero-form step functions} \\
&&& \theta(x^k- x^k_0) \\
&&&\\
\hline 
&&&\\
\text{zero-form}~ \lambda^k & & \lambda^k \rightarrow \lambda^k + 2\pi m^k + \nu  & \text{zero-form step functions} \\
&&& \theta(x^k- x^k_0) \\
&&&\\
\hline 
&&&\\
(0+1)\text{-form}~ \tilde{\beta}^k & \tilde{\beta}^k \wedge e^k = 0 & \tilde{\beta}^k \rightarrow \tilde{\beta}^k -2\pi d m^k & \text{one-form delta functions} \\
&&& \delta(x^k- x^k_0)\, d x^k \\
&&&\\
\hline 
&&&\\
\text{one-form}~ \mu &  & \mu \rightarrow \mu + d\nu & \text{zero-form step functions} \\
&&&\\
\hline 
&&&\\
\text{$x^k$-dependent}  &  && \text{zero-form step functions} \\
\text{function}~ m^k \in  \Z &&& \theta(x^k- x^k_0)  \\
&&&\\
\hline \end{array}\right. 
\end{align*}
\caption{The foliated $B$-type one-form gauge field and its gauge parameters.}
\label{3tableb}
\end{table}

\subsubsection{3+1d Foliated $BF$ Lagrangian}

The foliated $BF$ Lagrangian %
is similar to (\ref{flagrangian}), where $B^k$ is a $B$-type one-form foliated gauge field and $b$ is a two-form gauge field. In the special case where the foliations are flat, $n_f = 3$ (i.e., $e^k = dx^k\ (k = 1,2,3)$), $M_k=N$ and $n_k = 1$, the foliated Lagrangian can be written as
\begin{align}
    \L_f = \sum^{3}_{k=1} \frac{iN}{2\pi} (dB^k + b)\wedge A^k \wedge dx^k + \frac{iN}{2\pi} b \wedge da\,. \label{31folilag}
\end{align}

The equations of motion are
\begin{subequations}
\begin{align}
     \frac{N}{2\pi}(dB^k + b) \wedge dx^k = 0\,,  \label{3feom1}\\
     \frac{N}{2\pi}db = 0\,, \label{3feom2}\\
     \frac{N}{2\pi}dA^k \wedge dx^k = 0\,, \label{3feom3}\\
     \sum^{3}_{k=1} \frac{N}{2\pi} A^k \wedge dx^k + \frac{N}{2\pi} da =0\,. \label{3feom4}
\end{align}
\end{subequations}

The gauge transformations are
 \begin{subequations}
 \begin{align}
     A^k \wedge dx^k &\rightarrow A^k \wedge dx^k + d\zeta^k \wedge dx^k\,, \label{3fgauge1} \\
     B^k &\rightarrow B^k   + d\lambda^k + \beta^k dx^k - \mu\,,  \label{3fgauge2} \\
     a &\rightarrow a + d\lambda - \sum^{3}_{k=1} \zeta^k dx^k\,, \label{3fgauge3} \\
     b &\rightarrow b + d\mu\,, \label{3fgauge4}
 \end{align}
 \end{subequations}
where $\zeta^k$, $\lambda^k$, $\beta^k dx^k = \tilde{\beta}^k$ and $\mu$ are the gauge parameters explained in Section \ref{ffield} and Section \ref{3ffield}, and $\lambda$ are zero-form bulk gauge parameters that also have their own gauge transformations. The gauge transformation of $\lambda$ is $\lambda \rightarrow \lambda + 2\pi \xi^1 + 2\pi \xi^2 + 2\pi \xi^3$, where $\xi^k$ are $x^k$-dependent functions valued in integers explained in Section \ref{ffield}. As in the case of $2+1$ dimensions, the bulk fields $a$, $b$ and their gauge parameters can have singularities and discontinuities.

Integrating the fields out and considering specific field configurations, we can show that the following quantities are quantized:
\begin{subequations}
    \begin{align}
    \oint_{C_1^0} a \in \frac{2\pi}{N}\Z\,, \label{3aqua} \\ 
    \oint_{S_2^k} A^k \wedge dx^k \in  \frac{2\pi}{N}\Z\,, \label{3akqua} \\
    \oint_{S_2} b \in  \frac{2\pi}{N}\Z\,, \label{3bqua} \\
    \oint_{C_1^q} \sum_{k=1}^3 q_k B^k \in \frac{2\pi}{N}\Z\,, \label{3bkqua}
\end{align}
\end{subequations}
where $C_1^0$ is a closed one-dimensional loop along the time $x^0$ direction, $S_2$ is an arbitrary closed two-dimensional surface, and $S_2^k$ is a two-dimensional strip with a fixed width along the $x^k$ direction. The charges $q_k$ are  integers that satisfy $\sum_{k=1}^3 q_k =0$ and $C_1^q$ is a one-dimensional loop supported on the intersection of leaves with nonzero $q_k$.

\subsubsection{Gauge-Invariant Operators}

Let us discuss the gauge-invariant operators, which describe excitations moving in spacetime.

The first one is
\begin{align}
    F^q[C_1^0] = \exp \left[ i q\oint_{C_1^0} a \right]\,, \label{3ffracton}
\end{align}
$F^q[C_1^0]$ is a $\Z_N$ operator: $F^{q+N} =  F^q$. Since $C_1^0$ is a line in the time direction, this one-dimensional operator is the defect operator that describes a fracton, which cannot move in space.

The second one is
\begin{align}
    L[C_1^q] = \exp \left[ i \oint_{C_1^q} \sum_{k=1}^3 q_k B^k \right] \label{flineon}\,,
\end{align}
where $q = (q_k)_k$ is a charge vector where the components are  integers satisfying $\sum_{k=1}^3 q_k =0$, and $C_1^q$ is a closed one-dimensional line in an intersection of leaves, each of which is a single leaf for foliation $k$ with $q_k \neq 0$. From (\ref{3bkqua}), we can see that $L[C_1^{(N q)}] = 1$. Any charge vector $q$ can be spanned by $q^1 = (0,1,-1)$ and $q^2 = (-1,0,1)$, and we define $q^3 = -q^1 - q^2 = (1,-1,0)$, and then the corresponding operators are
\begin{subequations}
\begin{align}
     L_1^q[C_1^{01}] &= \exp \left[ i q \oint_{C_1^{01}} ( B^2 - B^3 )\right] \label{fxlineon}\,, \\
     L_2^q[C_1^{02}] &= \exp \left[ i q \oint_{C_1^{02}} ( B^3 - B^1 ) \right] \label{fylineon}\,, \\
     L_3^q[C_1^{03}] &= \exp \left[ i q \oint_{C_1^{03}} ( B^1 - B^2 ) \right] \label{fzlineon}\,,
\end{align}
\end{subequations}
where $C_1^{0k}$ is a closed one-dimensional loop in the $(x^0,x^k)$-plane. The one-dimensional operator $L_k^q[C_1^{0k}]$ describes a $x^k$-lineon that can only move along the $x^k$ direction. Therefore $L[C_1^q]$ is written as a product of the lineon operators. From (\ref{3bkqua}), we can see that $L_k^q[C_1^{0k}]$ are $\Z_N$ operators: $L_k^{q+N} = L_k^q$. The line operators $L_k^q[C_1^k]$ are the symmetry operators that generate $\Z_N$ tensor global symmetries, which are subsystem symmetries.

The third ones are 
\begin{align}
    W_k^q[S^k_2] = \exp \left[ i q \oint_{S_2^k} A^k \wedge dx^k \right] \,, \quad k=1,2,3\,.  \label{3fstrip}
\end{align}
Similarly, $W_k^q[S^k_2]$ are $\Z_N$ operators: $W^{q+N} = W^q$. These two-dimensional strip operators describe a dipole of fractons separated along the $x^k$ direction, which can move in the other directions in space, like a planon. If $S^k_2$ are in the space, these operators become the symmetry operators that generate $\Z_N$ dipole global symmetries, which are also subsystem symmetries.

These two types of symmetry operators are the charged objects under the other symmetry. That is,
$L^p$ and $W_k^q[S^k_2]$ satisfy the following relations at equal time:
\begin{subequations}
\begin{align}
    L_2^p[C_1^2] \ W_1^q[S^1_2] =  \mathrm{e}^{2\pi i p q I(C_1^2,S^1_2) /N} \  W_1^q[S^1_2]  \ L_2^p[C_1^2]\,, \quad \text{if} \quad x^1_1 < x^1 < x^1_2\,, \\
    L_3^p[C_1^3] \ W_1^q[S^1_2] =  \mathrm{e}^{2\pi i p q I(C_1^3,S^1_2) /N} \  W_1^q[S^1_2]  \ L_3^p[C_1^3]\,, \quad \text{if} \quad x^1_1 < x^1 < x^1_2\,,
\end{align}
\end{subequations}
where $S^1_2 = [x^1_2,x^1_1] \times C^{23}_1$ , and $I$ is the intersection number. Similar relations holds in the other directions.

The forth one is
\begin{align}
    K^q_{12} [C_1^{03} \times \mathcal{C}^{12}_1] = \exp \left[ i q \oint_{C_1^{03} \times \mathcal{C}^{12}_1} (A^1\wedge dx^1 + A^2\wedge dx^2 )  \right] \,, \label{3fdipfra12}
\end{align}
where $\mathcal{C}^{12}_1$ is a one-dimensional line connecting $(x^1_1,x^2_1)$ to $(x^1_2,x^2_2)$ in the $(x^1,x^2)$-plane. $K^q_{12}$ is the strip operator that describes a dipole of fractons at $(x^1_1,x^2_1,x^3)$ and $(x^1_2,x^2_2,x^3)$, which can move in the $x^3$ direction, like a $x^3$-lineon. Using the Stokes' theorem, the equations of motion \eqref{3feom3}, $\mathcal{C}^{12}_1$ can be deformed to $[x^1_1,x^1_2]\times \{x^2_1\} + \{x^1_2\} \times [x^2_1,x^2_2]$, and this special case we write $\tilde{K}^q_3$ as
\begin{equation}
\begin{split}
    K^q_{12}[C_1^{03} \times \mathcal{C}^{12}_1] =  &\exp \left[ i q \oint_{C_1^{03} \times [x^1_1,x^1_2]\times \{x^2_1\}} A^1\wedge dx^1 \right] \\ &\quad \times \exp \left[ i q \oint_{C_1^{03} \times \{x^1_2\} \times [x^2_1,x^2_2]}  A^2\wedge dx^2  \right] \\
    = &W^q_1\left[C_1^{03} \times [x^1_1,x^1_2]\times \{x^2_1\}\right] W^q_2\left[C_1^{03} \times \{x^1_2\} \times [x^2_1,x^2_2]\right] \,.
\end{split}
\end{equation}
Similarly, we have the strip operators $K^q_{23}$ and $K^q_{31}$. 

In addition, the bulk $3+1$d $BF$ theory has a gauge-invariant operator
\begin{align}
    T^q[S_2] = \exp \left[ i q \oint_{S_2} b\right]\,. \label{3bop}
\end{align}
From (\ref{3bqua}), this surface operator is also a $\Z_N$ operator: $T^{q+N} = T^q$. This operator has the winding action on the gauge-invariant operator (\ref{3ffracton}) as
\begin{align}
 T^p[S_2]\cdot  F^q[C_1^0] = \mathrm{e}^{-2\pi i p q /N} F^q[C_1^0]\,,
\end{align}
when $S_2$ surrounds $C_1^0$. 
Without the defect operator $F^q$, the $b$ operator $T^q$ becomes trivial, which corresponds to a time-like symmetry \cite{Gorantla:2022eem}.
When $S_2$ is $S_2^{123,\text{cube}}$ that is the surface of $[x^1_1,x^1_2]\times [x^2_1,x^2_2] \times [x^3_1,x^3_2]$, using the equation of motion \eqref{3feom1}, 
a part of the integral in the definition of $T^q$ can be performed as 
\begin{equation}
\begin{split}
    T^q[S_2^{123,\text{cube}}] &= \exp \bigg[ i q \oint_{S_2^{123,\text{cube}}} \left\{ -(\partial_1 B^3_2 + \partial_2 B^3_1 ) dx^1 dx^2 - (\partial_2 B^1_3 + \partial_3 B^1_2 ) dx^2 dx^3 \right. \\
    & \hspace{7cm} \left. -(\partial_3 B^2_1 + \partial_1 B^2_3 ) dx^3 dx^1 \right\} \bigg] \\
    & = \exp \bigg[- i q \int_{x^1_1}^{x^1_2} \left\{ \Delta_{23} (B^2_1 - B^3_1)(x^2_1,x^2_2,x^3_1,x^3_2) \right\} dx^1 \bigg] \\
    & \qquad \times  \exp \bigg[- i q \int_{x^2_1}^{x^2_2} \left\{ \Delta_{31} (B^3_2 - B^1_2)(x^3_1,x^3_2,x^1_1,x^1_2) \right\} dx^2 \bigg] \\
    & \qquad \quad \times  \exp \bigg[- i q \int_{x^3_1}^{x^3_2} \left\{ \Delta_{12} (B^1_3 - B^2_3)(x^1_1,x^1_2,x^2_1,x^2_2) \right\} dx^3 \bigg] \,. \label{3ftimelike}
\end{split}
\end{equation}
This cage operator is localized on the edges of the rectangular cuboid whose surface is $S_2^{123,\text{cube}}$. 
When $S_2$ is $S_2^{012,\text{cube}}(x^1_1,x^1_2,x^2_1,x^2_2)$ that is  $C_1^0 \times C_1^{12,\text{rect}}(x^1_1,x^1_2,x^2_1,x^2_2)$ extended along the $x^0$ direction, the $b$ operator can be written as
\begin{equation}
\begin{split}
    T^q[S_2^{012,\text{cube}}(x^1_1,x^1_2,x^2_1,x^2_2)] &= \exp \bigg[ i q \oint_{S_2^{012,\text{cube}}} \left\{ -(\partial_0 B^2_1 + \partial_1 B^2_0 ) dx^0 dx^1 \right. \\ 
    & \hspace{3cm} - \left. (\partial_2 B^1_0 + \partial_0 B^1_2 ) dx^2 dx^0 \right\}  \bigg] \\
    & = \exp \left[- i q \oint_{C_1^0} \Delta_{12}  (B^1_0 - B^2_0)(x^1_1,x^1_2,x^2_1,x^2_2)  dx^0 \right] \,. 
\end{split}
\end{equation}
This operator describes a quadrupole of $x^3$-lineons $L^q_3[C_1^0]$ at the corners of the rectangle $C_1^{12,\text{rect}}(x^1_1,x^1_2,x^2_1,x^2_2)$, which is trivial. Similarly, we also have the operators that describe a quadrupole of $x^1$-lineons and a quadrupole of $x^2$-lineons.

As opposed to the $(2+1)$-dimensional case, the time-like $T^q$ operator does correspond to physical excitations that are the lineons sitting at the corners of the time-slice of the operator. However the operator is trivial as it cannot be detected by a time-like symmetry.\footnote{A lineon can be detected by a time-like symmetry generated by belt operator, which will be introduced later. However there is no way to place the belt operator so that it detects the quadrupole $T^q[S^{012,\text{cube}}_2]$ without intersecting the belt operator with $S^{012,\text{cube}}_2$. Thus the operator $T^q[S^{012,\text{cube}}_2]$ is trivial.}  This is also different from the case of an ordinary $BF$ theory, in which a $(1+1)$-dimensional operator like $T^q$ corresponds to a one-dimensional, or string, excitation.

\subsection{3+1d Exotic $BF$ Theory} \label{31ebf}

We will review the exotic $BF$ theory in $3+1$ dimensions\cite{Seiberg:2020cxy}.

\subsubsection{Tensor Gauge Fields}

In $3+1$ dimensions, the $90$ degree rotations generate the orientation-preserving cubic group, which is isomorphic to the permutation group $S_4$. Then each tensor gauge field is in a representation of $S_4$. The irreducible representations of $S_4$ are classified as the following tensors (see Appendix in \cite{Seiberg:2020wsg,Seiberg:2020cxy}):
\begin{equation}
\begin{split}
    \bm{1} \, &: \quad S\,, \\
    \bm{1'}  &: \quad T_{(ijk)} \,,\quad i \neq j \neq k \,, \\
    \bm{2}  \, &: \quad B_{[ij]k} \,,\quad i \neq j \neq k \,, \quad B_{[ij]k} + B_{[jk]i} + B_{[ki]j} = 0 \,,\\
            &\quad \quad B_{i(jk)} \,,\quad i \neq j \neq k \,, 
    \quad B_{i(jk)} + B_{j(ki)} + B_{k(ij)} = 0 \,,        \\
    \bm{3} \, &: \quad  V_i \,, \\
    \bm{3'}   &: \quad E_{ij} \,,\quad i \neq j \,, 
    \quad E_{ij} = E_{ji} \,,\\
\end{split}
\end{equation}
where $i$, $j$, $k$ are $1$ or $2$ or $3$, the indices $[ij]$ are antisymmetrized, the indices $(ij)$ are symmetrized. The two bases of irreducible representation $\bm{2}$ are related as
\begin{subequations}
\begin{align}
    B_{i(jk)} &= B_{[ij]k} + B_{[ik]j} \,,   \label{rep2rel1} \\
    B_{[ij]k} &= \frac{1}{3} ( B_{i(jk)} + B_{j(ik)}) \,. \label{rep2rel2}
\end{align}
\end{subequations}
In the following, sums are taken over $1$, $2$ and $3$; for example,  $\sum_{i,j,k} \hat{A}^{k(ij)}B_{k(ij)} = 2 \hat{A}^{1(23)}B_{1(23)} + 2 \hat{A}^{2(31)}B_{2(31)} + 2 \hat{A}^{3(12)}B_{3(12)}$.

The exotic $BF$ theory contains two $U(1)$ tensor gauge fields $(A_0,A_{ij})$ in $(\bm{1},\bm{3'})$ and $(\hat{A}_0^{i(jk)},\hat{A}^{ij})$ in $(\bm{2},\bm{3}')$ as its fields. The gauge transformation of $(A_0,A_{ij})$ is
\begin{subequations}
\begin{align}
    A_0 &\rightarrow A_0 + \partial_0 \alpha \,, \label{3egauge1} \\
    A_{ij} &\rightarrow A_{ij} + \partial_i \partial_j \alpha\,, \label{3egauge2}
\end{align}
\end{subequations}
where $\alpha$ is a $U(1)$-valued gauge parameter: $\alpha \sim \alpha + 2\pi$, in $\bm{1}$. The gauge parameter $\alpha$ has its own gauge transformation: $\alpha \rightarrow \alpha + 2\pi \tilde{n}^1 + 2\pi \tilde{n}^2 + 2\pi \tilde{n}^3$, where $\tilde{n}^k$ are $x^k$-dependent functions valued in integers. $A_{ij}$ can have delta function singularities and have step function discontinuities, while $A_0$, $\alpha$ can have step function discontinuities as in Table~\ref{3exotictable}, where $\tilde{f}_0^k$, $\tilde{f}_{ij}^k$, $\tilde{g}^k$, $\tilde{h}_0^{i(jk),l}$, $\tilde{h}^{ij,l}$ and $\tilde{s}^{i(jk),l}$ are some continuous functions  with appropriate periodicity conditions.\footnote{Functions $f(x;\hat{x^k})$ mean $x^k$-independent functions.} 
The gauge parameter $\tilde{n}^k$ can have step function discontinuities in the $x^k$ direction. The gauge-invariant electric and magnetic fields of $(A_0,A_{ij})$ are 
\begin{gather}
    E_{ij} = \partial_0 A_{ij} - \partial_i \partial_j A_0 \,, \\
    B_{[ij]k} = \partial_i A_{jk} - \partial_j A_{ik} \quad \text{or}\quad  B_{k(ij)} = 2\partial_k A_{ij} - \partial_i A_{kj} - \partial_j A_{ki} \,.
\end{gather}
The gauge transformation of the other field  $(\hat{A}_0^{i(jk)},\hat{A}^{ij})$ is
\begin{subequations}
\begin{align}
    \hat{A}_0^{i(jk)} &\rightarrow \hat{A}_0^{i(jk)} + \partial_0   \hat{\alpha}^{i(jk)} \,, \label{3egauge3} \\
    \hat{A}^{ij} &\rightarrow \hat{A}^{ij} + \partial_k \hat{\alpha}^{k(ij)}  \label{3egauge4}\,,
\end{align}
\end{subequations}
where $\hat{\alpha}^{i(jk)}$ is a $U(1)$-valued gauge parameter in $\bm{2}$. The gauge parameter $\hat{\alpha}^{i(jk)}$ has its own gauge transformation:  $\hat{\alpha}^{i(jk)} \rightarrow \hat{\alpha}^{i(jk)} + 2\pi \tilde{m}^j - 2\pi \tilde{m}^k$, where $\tilde{m}^k$ are $x^k$-dependent functions valued in integers. $\hat{A}_0^{i(jk)}$, $\hat{A}^{jk}$ and $\hat{\alpha}^{i(jk)}$ can have step function discontinuities in the $x^j$ and $x^k$ directions as in Table~\ref{3exotictable}. The gauge parameter $\tilde{m}^k$ can have step function discontinuities in the $x^k$ direction.
The gauge-invariant electric and magnetic fields of $(\hat{A}_0^{i(jk)},\hat{A}^{ij})$ are 
\begin{align}
    &\hat{E}^{ij} = \partial_0 \hat{A}^{ij} - \partial_k \hat{A}^{k(ij)}_0 \,, \\
     &\hat{B} = \frac{1}{2} \sum_{i,j} \partial_i \partial_j \hat{A}^{ij} \,.
\end{align}

\begin{table}[htbp]
\centering
\begin{align*}
\left. \begin{array}{|c|c|c|}
\hline 
&&\\
 \text{Gauge fields} & \text{Gauge transformation} & \text{Terms including} \\
 \text{and parameters} && \text{singularities and discontinuities} \\
&& \\
\hline 
&& \\
 A_0 &   A_0 \rightarrow A_0 +\partial_0 \alpha  & \sum_{k=1}^3 \tilde{f}_0^k (x;\hat{x^k})\theta(x^k-x^k_0)  \\
&& \\ 
\hline 
&& \\
 A_{ij} &   A_{ij} \rightarrow A_{ij} +\partial_i \partial_j \alpha   &  \tilde{f}_{ij}^i(x;\hat{x^i})\delta(x^i-x^i_0) +  \tilde{f}_{ij}^j(x;\hat{x^j})\delta(x^j-x^j_0)  \\
&& + \tilde{f}_{ij}^k(x;\hat{x^k})\theta(x^j-x^j_0) \quad (k\neq 0,i,j) \\
&& \\
\hline 
&& \\
 \alpha &  \alpha \rightarrow \alpha + 2\pi \tilde{n}^1 + 2\pi \tilde{n}^2 + 2\pi \tilde{n}^3  & \sum_{k=1}^3 \tilde{g}^k (x;\hat{x^k})\theta(x^k-x^k_0)  \\
&& \\ 
\hline
&& \\
 \hat{A}_0^{i(jk)} &  \hat{A}_0^{i(jk)} \rightarrow \hat{A}_0^{i(jk)} + \partial_0 \hat{\alpha}^{i(jk)}  &   \tilde{h}_0^{i(jk),j}(x;\hat{x^j})\theta(x^j-x^j_0)  \\
&& \qquad + \tilde{h}_0^{i(jk),k}(x;\hat{x^k})\theta(x^k-x^k_0)\\ 
&& \\
\hline
&& \\
\hat{A}^{ij} & \hat{A}^{ij} \rightarrow \hat{A}^{ij} + \partial_k \hat{\alpha}^{k(ij)} & \tilde{h}^{ij,i}(x;\hat{x^i})\theta(x^i-x^i_0) +  \tilde{h}^{ij,j}(x;\hat{x^j})\theta(x^j-x^j_0)\\
&& \\
\hline
&& \\
\hat{\alpha}^{i(jk)} & \hat{\alpha}^{i(jk)} \rightarrow \hat{\alpha}^{i(jk)} + 2\pi \tilde{m}^j - 2\pi \tilde{m}^k & \tilde{s}^{i(jk),j}(x;\hat{x^j})\theta(x^j-x^j_0)  \\ 
&& \qquad + \tilde{s}^{i(jk),k}(x;\hat{x^k})\theta(x^k-x^k_0) \\
&& \\
\hline
\end{array}\right. 
\end{align*}
\caption{Singularities and discontinuities of the tensor gauge fields and their gauge parameters.}
\label{3exotictable}
\end{table}

\subsubsection{3+1d Exotic $BF$ Lagrangian}

The exotic $BF$ Lagrangian is
\begin{align}
    \L_e = \frac{i N}{2\pi}\left( \frac{1}{2} \sum_{i,j} A_{ij} \hat{E}^{ij} + A_0 \hat{B} \right) \,. 
\end{align}
Integrating by parts, and using $\sum_{i,j,k}(\partial_i A_{jk} + \partial_j A_{ki} + \partial_k A_{ij}) \hat{A}_0^{k(ij)} = \sum_{i,j,k} \partial_i A_{jk} ( \hat{A}_0^{k(ij)} + \hat{A}_0^{i(jk)} + \hat{A}_0^{j(ki)}) = 0$ and (\ref{rep2rel1}), the Lagrangian can be written as
\begin{align}
    \L_e = \frac{i N}{2\pi} \left(  \frac{1}{6}\sum_{i,j,k}  \hat{A}_0^{k(ij)} B_{k(ij)} + \frac{1}{2}\sum_{i,j} \hat{A}^{ij} E_{ij} \right) \,. 
\end{align}

The equations of motion are
\begin{subequations}
\begin{align}
    &\frac{N}{2\pi} E_{ij} = 0 \,,  \label{3eeom1} \\
    &\frac{N}{2\pi} B_{k(ij)} = 0 \,,  \label{3eeom2} \\
    &\frac{N}{2\pi} \hat{E}^{ij} = 0 \,,  \label{3eeom3} \\
    &\frac{N}{2\pi} \hat{B} = 0 \,.  \label{3eeom4}
\end{align}
\end{subequations}

Integrating specific fields out, we can show that the following quantities are quantized.
\begin{subequations}
\begin{gather}
    \oint_{C_1^0} dx^0 A_0 \in \frac{2\pi}{N}\Z \,, \label{3eaqua} \\ 
    \oint_{S^1_2} ( dx^0 dx^1 \partial_1 A_0 + dx^2 dx^1 A_{12} + dx^3 dx^1 A_{31} )  \in  \frac{2\pi}{N}\Z \,, \label{3eakqua1} \\
    \oint_{S^2_2} ( dx^0 dx^2 \partial_2 A_0 + dx^1 dx^2 A_{12} + dx^3 dx^2 A_{23}) \in  \frac{2\pi}{N}\Z \,, \label{3eakqua2} \\
    \oint_{S^3_2} ( dx^0 dx^3 \partial_3 A_0 + dx^1 dx^3 A_{31} + dx^2 dx^3 A_{23} )  \in  \frac{2\pi}{N}\Z \,, \label{3eakqua3} \\
    \oint_{C_1^{0k}} ( dx^0 \hat{A}_0^{k(ij)} + dx^k \hat{A}^{ij}) \in \frac{2\pi}{N}\Z\,, \label{3hatqua}
\end{gather}
\end{subequations}
where $C_1^0$ is a closed one-dimensional loop along the time $x^0$ direction, $S^k_2$ is a two-dimensional strip with a fixed width along the $x^k$ direction, and $C_1^{01}$ is a closed one-dimensional loop in the $(x^0,x^k)$-plane.

\subsubsection{Gauge-Invariant Operators}

Let us discuss gauge-invariant operators. The defect operator that describes fractons is
\begin{align}
    \tilde{F}^q[C_1^0] = \exp \left[ i q \oint_{C_1^0} dx^0 A_0\right] \,. \label{3efracton}
\end{align}
As in the case of the foliated $BF$ theory, the deformation of $C^0_1$ would break the gauge invariance of the operator.

The defect operators that describe lineons are
\begin{align}
    \tilde{L}^q_k[C_1^{0k}] = \exp \left[ i q \oint_{C_1^{0k}} ( dx^0 \hat{A}_0^{k(ij)} + dx^k \hat{A}^{ij}) \right] \,, \quad k=1,2,3 \,. \label{3elineon}
\end{align}
If $C_1^{0k}$ are along the $x^k$ direction, $\tilde{L}^q_k$ are the symmetry operators that generate $\Z_N$ tensor global symmetries.

The strip operators that describe a dipole of fractons separated along the $x^k$ directions are
\begin{subequations}
\begin{align}
    \tilde{W}_1^q\left[ S^1_2 \right] = \exp \left[ i q \oint_{S^1_2} ( dx^0 dx^1 \partial_1 A_0 + dx^2 dx^1 A_{12} + dx^3 dx^1 A_{31} )  \right] \,, \label{3estrip1}\\
    \tilde{W}_2^q\left[ S^2_2  \right] = \exp \left[ i q \oint_{S^2_2} ( dx^0 dx^2 \partial_2 A_0 + dx^1 dx^2 A_{12} + dx^3 dx^2 A_{23})   \right]\,, \label{3estrip2}\\
    \tilde{W}_3^q\left[ S^3_2  \right] = \exp \left[ i q \oint_{S^3_2} ( dx^0 dx^3 \partial_3 A_0 + dx^1 dx^3 A_{31} + dx^2 dx^3 A_{23} )   \right]\,. \label{3estrip3}
\end{align}
\end{subequations}
If $S^k_2$ are in the $(x^1,x^2,x^3)$-plane, $\tilde{W}_k^q$ are the symmetry operators that generate $\Z_N$ dipole global symmetries.

The two types of symmetry operators satisfy the following relations at equal time:
\begin{subequations}
\begin{align}
    \tilde{L}_2^p[C_1^2] \ \tilde{W}_1^q[S^1_2] =  \mathrm{e}^{2\pi i p q I(C_1^2,S^1_2) /N} \  \tilde{W}_1^q[S^1_2]  \ \tilde{L}_2^p[C_1^2]\,, \quad \text{if} \quad x^1_1 < x^1 < x^1_2\,, \\
    \tilde{L}_3^p[C_1^3] \ \tilde{W}_1^q[S^1_2] =  \mathrm{e}^{2\pi i p q I(C_1^3,S^1_2) /N} \  \tilde{W}_1^q[S^1_2]  \ \tilde{L}_3^p[C_1^3]\,, \quad \text{if} \quad x^1_1 < x^1 < x^1_2\,,
\end{align}
\end{subequations}
where $S^1_2 = [x^1_2,x^1_1] \times C_1^{23}$, and $I$ is the intersection number. Similar relations holds in the other directions. These symmetries in the exotic $BF$ theory have the same structure as the foliated $BF$ theory discussed in Section \ref{31fbf}.

In addition, we can consider the strip operator that describes a dipole of fractons at $(x^1_1,x^2_1,x^3)$ and $(x^1_2,x^2_2,x^3)$, which can move in the $x^3$ direction, like a $x^3$-lineon:
\begin{align}
    \tilde{K}^q_{12}[C_1^{03} \times \mathcal{C}^{12}_1] = \exp \left[ i q \oint_{C_1^{03} \times \mathcal{C}^{12}_1} (dx^0 dx^1 \partial_1 A_0 + dx^0 dx^2 \partial_2 A_0 + dx^3 dx^1 A_{31} + dx^3 dx^2 A_{23})  \right] \,, \label{3dipfra}
\end{align}
where $\mathcal{C}^{12}_1$ is a one-dimensional line connecting $(x^1_1,x^2_1)$ to $(x^1_2,x^2_2)$ in the $(x^1,x^2)$-plane. Using the Stokes' theorem, the equations of motion \eqref{3eeom1}, and $\frac{N}{2\pi} B_{[ij]k} = 0$ from \eqref{3eeom2}, $\mathcal{C}^{12}_1$ can be deformed to $[x^1_1,x^1_2]\times \{x^2_1\} + \{x^1_2\} \times [x^2_1,x^2_2]$, and in this special case we write $\tilde{K}^q_{12}$ as
\begin{equation}
\begin{split}
    \tilde{K}^q_{12}[C_1^{03} \times \mathcal{C}^{12}_1] =  &\exp \left[ i q \oint_{C_1^{03} \times [x^1_1,x^1_2]\times \{x^2_1\}} (dx^0 dx^1 \partial_1 A_0 +  dx^3 dx^1 A_{31} )\right] \\ &\quad \times \exp \left[ i q \oint_{C_1^{03} \times \{x^1_2\} \times [x^2_1,x^2_2]}  (dx^0 dx^2 \partial_2 A_0 + dx^3 dx^2 A_{23})  \right] \\
    = &\tilde{W}^q_1\left[C_1^{03} \times [x^1_1,x^1_2]\times \{x^2_1\}\right] \tilde{W}^q_2\left[C_1^{03} \times \{x^1_2\} \times [x^2_1,x^2_2]\right] \,.
\end{split}
\end{equation}
Similarly, we have the strip operators $\tilde{K}^q_{23}$ and $\tilde{K}^q_{31}$. 

Also, we can consider the strip operators that describe a dipole of $x^1$-lineons and a dipole of $x^2$-lineons, separated along the $x^3$ direction, which can move in the other directions in space, like a planon:
\begin{align}
    \tilde{P}^q_{3,1}[S^3_2]  = \exp \left[ i q \oint_{S^3_2} \left( dx^0 dx^3 \partial_3 \hat{A}_0^{1(23)} + dx^1 dx^3 \partial_3 \hat{A}^{23} - dx^2 dx^3 ( \partial_3 \hat{A}^{31} + \partial_2 \hat{A}^{12} ) \right) \right]  \,, \label{3diplin1} \\
    \tilde{P}^q_{3,2}[S^3_2]  = \exp \left[ i q \oint_{S^3_2} \left( dx^0 dx^3 \partial_3 \hat{A}_0^{2(31)} + dx^2 dx^3 \partial_3 \hat{A}^{31} - dx^1 dx^3 ( \partial_3 \hat{A}^{23} + \partial_1 \hat{A}^{12} ) \right) \right]  \,. \label{3diplin2}
\end{align}
Note that 
\begin{equation}
\begin{split}
    \tilde{P}^q_{3,1} &= \{\tilde{P}^q_{3,2}\}^{-1} \exp \left[ -i q \oint_{S^3_2} ( dx^0 dx^3 \partial_3 \hat{A}_0^{3(12)} + dx^2 dx^3 \partial_2 \hat{A}^{12} + dx^1 dx^3  \partial_1 \hat{A}^{12} ) \right] \\ 
    &=  \{\tilde{P}^q_{3,2}\}^{-1} \{\tilde{P}^q_{3,3}\}^{-1} \,,  
\end{split}
\end{equation}
where $\tilde{P}^q_{3,3}$ represents a dipole of $x^3$-lineons separated along the $x^3$ direction, which is trivial. Using the Stokes' theorem and the equations of motion \eqref{3eeom3} and \eqref{3eeom4}, $S^3_2$ can be deformed to $C_1^{01}\times \{x^2\} \times[x^3_1,x^3_2]$, and in this special case we write $\tilde{P}^q_{3,1}[S^3_2]$ as
\begin{equation}
\begin{split}
     \tilde{P}^q_{3,1}[S^3_2] =  &\exp \left[ i q \oint_{C_1^{01}\times \{x^2\} \times[x^3_1,x^3_2]} ( dx^0 dx^3 \partial_3 \hat{A}_0^{1(23)} + dx^1 dx^3 \partial_3 \hat{A}^{23}  ) \right] \\
    = &\exp \left[ i q \oint_{C_1^{01} (x^2,x^3_2)} ( dx^0 \hat{A}_0^{1(23)} + dx^1  \hat{A}^{23} ) \right] \\
    &\quad \times \exp \left[ -i q \oint_{C_1^{01} (x^2 ,x^3_1)} ( dx^0 \hat{A}_0^{1(23)} + dx^1  \hat{A}^{23} ) \right] \\
    = &\tilde{L}^q_1\left[C_1^{01}(x^2,x^3_2) \right] \left\{\tilde{L}^q_1\left[C_1^{01}(x^2,x^3_1) \right]\right\}^{-1} \,,
\end{split}
\end{equation}
where $C^{01}_1(x^2,x^3)$ is a closed one-dimensional loop in the $(x^0,x^1)$-plane at $(x^2,x^3)$. Similarly, we have the strip operators $\tilde{P}^q_{k,i}$ for $(k,i) = (1,2),(1,3),(2,1),(2,3)$.

As in the case of $2+1$ dimensions, there is a gauge-invariant operator that can detect the fracton operator:
\begin{equation}
\begin{split}
    \tilde{T}^q\left[S_2^{123,\text{cube}}\right] 
    & = \exp \bigg[- i q \int_{x^1_1}^{x^1_2} dx^1 \left\{ \Delta_{23} \hat{A}^{23} (x^2_1,x^2_2,x^3_1,x^3_2) \right\}  \bigg] \\
    & \qquad \times  \exp \bigg[- i q \int_{x^2_1}^{x^2_2} dx^2 \left\{ \Delta_{31} \hat{A}^{31} (x^3_1,x^3_2,x^1_1,x^1_2) \right\}  \bigg] \\
    & \qquad \quad \times  \exp \bigg[- i q \int_{x^3_1}^{x^3_2} dx^3 \left\{ \Delta_{12} \hat{A}^{12}(x^1_1,x^1_2,x^2_1,x^2_2) \right\}  \bigg] \,. \label{3etimelike1}
\end{split}
\end{equation}
This cage operator is localized on the edges of the rectangular cuboid whose surface is $S_2^{123,\text{cube}}$. Without the defect operator $\tilde{F}^q$, $\tilde{T}^q$ becomes trivial, which corresponds to a time-like symmetry \cite{Gorantla:2022eem}. The operator $\tilde{T}^p$ can detect the fracton operator $\tilde{F}^q$:
\begin{align}
    \tilde{T}^p\left[S_2^{123,\text{cube}}\right] \cdot \tilde{F}^q[C^0_1] = \mathrm{e}^{-2\pi i p q/N} \tilde{F}^q[C^0_1] \,,
\end{align}
when $S_2^{123,\text{cube}}$ surrounds $C^0_1$.

This theory has another time-like symmetry whose operator can detect the lineons: the belt operator
\begin{equation}
\begin{split}
    \tilde{U}_{[12]3}^q\left[S_2^{3,\text{belt}}\right] &= \exp \left[ i q \int_{x^2_1}^{x^2_2} \int_{x^3_1}^{x^3_2} dx^2 dx^3 \left(A_{23}(x^1_2) - A_{23}(x^1_1) \right) \right] \\
    & \quad \quad  \times \exp \left[- i q \int_{x^1_1}^{x^1_2} \int_{x^3_1}^{x^3_2} dx^1 dx^3 \left(A_{31}(x^2_2) - A_{31}(x^2_1) \right) \right] \,, \label{3etimelike2}
\end{split}
\end{equation}
where $S_2^{3,\text{belt}}$ is $C_1^{12,\text{rect}} \times [x_1^3,x^3_2]$. Similarly, we also have $\tilde{U}_{[31]2}^q\left[S_2^{2,\text{belt}}\right]$ and $\tilde{U}_{[23]1}^q\left[S_2^{1,\text{belt}}\right]$. They act on the lineon operator as
\begin{subequations}
\begin{align}
    \tilde{U}_{[12]3}^p\left[S_2^{3,\text{belt}}\right] \cdot \tilde{L}_3^q[C_1^0] &= \tilde{L}_3^q[C_1^0]  \,, \\
    \tilde{U}_{[31]2}^p\left[S_2^{2,\text{belt}}\right] \cdot \tilde{L}_3^q[C_1^0] &= \mathrm{e}^{2\pi i p q /N}
    \tilde{L}_3^q[C_1^0] \,, \\
    \tilde{U}_{[23]1}^p\left[S_2^{1,\text{belt}}\right] \cdot \tilde{L}_3^q[C_1^0] &= \mathrm{e}^{-2\pi i p q /N}
    \tilde{L}_3^q[C_1^0] \,, 
\end{align}
\end{subequations}
where $S_2^{123,\text{cube}}$ that is the union of $S_2^{k,\text{belt}}$ surrounds $C_1^0$. Similar relations also hold for $\tilde{L}_1^q$ and $\tilde{L}_2^q$.

These gauge-invariant operators are $\Z_N$ operators: $q$ is an element of $\Z_N$.

\subsection{Correspondences in 3+1 Dimensions} \label{31corr}

As in the case of the $2+1$d version, the $3+1$d foliated $BF$ theory explained in Section \ref{31fbf} and the $3+1$d exotic $BF$ theory explained in Section \ref{31ebf} are equivalent in case that $e^k = dx^k$, $M_k=N$ and $n_k = 1$ $(k = 1,2,3)$. We identify the gauge-invariant operators in the foliated $BF$ theory with those of the exotic $BF$ theory. By matching the gauge-invariant operators, we can derive the correspondences of the gauge fields and parameters. 

First, let us consider the fracton defect operators. We identify the operators $F^q[C_1^0]$ with $\tilde{F}^q[C_1^0]$ defined in \eqref{3ffracton} and \eqref{3efracton}:
\begin{align}
    \exp \left[ i q\oint_{C_1^0} a \right] \simeq \exp \left[ i q \oint_{C_1^0} dx^0 A_0\right] \,,
\end{align}
which leads to the field correspondence
\begin{align}
    a_0 \simeq A_0 \,.
\end{align}
The gauge transformations of $a_0$ and $A_0$ explained in (\ref{3fgauge3}) and (\ref{3egauge1}) are
\begin{subequations}
    \begin{align}
        a_0 &\rightarrow a_0 + \partial_0 \lambda \,, \\
        A_0 &\rightarrow A_0 + \partial_0 \alpha \,,
    \end{align}
\end{subequations}
from which we obtain the gauge parameter correspondence
\begin{align}
\lambda \simeq \alpha \,. \label{3gcorr1}
\end{align}
Moreover, the gauge transformations of $\lambda$ and $\alpha$ are 
\begin{subequations}
    \begin{align}
        \lambda &\rightarrow \lambda+ 2\pi\xi^1 + 2\pi\xi^2 + 2\pi\xi^3 \,, \\
        \alpha &\rightarrow \alpha + 2\pi \tilde{n}^1 + 2\pi \tilde{n}^2 + 2\pi \tilde{n}^3 \,,
    \end{align}
\end{subequations}
which can be matched by
\begin{align}
         \xi^k \simeq \tilde{n}^k  \,.
\end{align}
In these correspondences, one can check that their singularities and discontinuities are also matched.

The equations of motion (\ref{3feom4}) in components are 
\begin{subequations}
\begin{align}
 \frac{N}{2\pi} ( A^k_0 + \partial_0 a_k - \partial_k a_0 ) &= 0 \,, \quad k=1,2,3 \,, \\
 \frac{N}{2\pi} ( A^k_i - A^i_k  + \partial_i a_k - \partial_k a_i ) &= 0 \,, \quad (k,i) = (1,2), (2,3), (3,1)\,. \label{3beomc3}
\end{align}
\end{subequations}
These equations of motion imply 
\begin{align}
    A^k_0 + \partial_0 a_k \simeq \partial_k A_0 \,, \quad k=1,2,3 \,. \label{3corr2}
\end{align}
Note that the gauge transformations by $\zeta^k$ cancel out.

Next, let us consider the strip operators. As in the case of $2+1$ dimensions, we define the modified gauge-invariant strip operators $W^q_{k,\text{mod}}[S^k_2]$ as
\begin{align}
    W^q_{k,\text{mod}}[S^k_2] = \exp \left[ i q \oint_{S^k_2} \left( A^k  \wedge dx^k + d(a_k dx^k) \right) \right] \,, \quad k=1,2,3 \,. \label{3fstripm}
\end{align}
We identify the operators $W^q_{k,\text{mod}}[S^k_2]$ with $\tilde{W}^q_k[S^k_2]$ defined in (\ref{3fstripm}), (\ref{3estrip1}), (\ref{3estrip2}) and (\ref{3estrip3}):
\begin{equation}
\begin{split}
    \exp \left[ i q \oint_{S^k_2} \left( A^k  \wedge dx^k + d(a_k dx^k) \right) \right] &\simeq 
    \exp \left[ i q \oint_{S^k_2} ( dx^0 dx^k \partial_k A_0 + dx^i dx^k A_{ik} + dx^j dx^k A_{jk} )  \right] \,, \\ (k,i,j) &= (1,2,3),(2,3,1),(3,1,2) \,,
\end{split}
\end{equation}
which lead to the field correspondences
\begin{align} 
    A^k_i + \partial_i a_k \simeq A_{ki} \,, \quad k\neq i \,, \quad k,i \in \{1,2,3\} \,, \label{3corr3} 
\end{align}
and also (\ref{3corr2}) again. The terms $\partial_i a_k$ make the gauge transformations match with those of $A_{ki}$ under the gauge parameter correspondence (\ref{3gcorr1}). Using the correspondences \eqref{3corr2} and \eqref{3corr3}, we find the correspondence of gauge-invariant operators between $K^q_{12}$ and $\tilde{K}^q_{12}$ defined in \eqref{3fdipfra12} and \eqref{3dipfra}. To be precise, as we did for $W^q_k$, the operator $K^q_{12}$ in the foliated side has to be modified as
\begin{align}
    K^q_{12,\text{mod}}[C_1^{03} \times \mathcal{C}^{12}_1] = \exp \left[ i q \oint_{C_1^{03} \times \mathcal{C}^{12}_1} \left(A^1\wedge dx^1 + A^2\wedge dx^2 + d(a_1 dx^1 + a_2 dx^2) \right)  \right] \,.
\end{align}

Lastly, let us consider the lineon operators. We identify $L_k^q[C^k_{0k}]$ with $\tilde{L}_k^q[C^k_1{0k}]$ defined in (\ref{fxlineon})-(\ref{fzlineon}) and (\ref{3elineon}):
\begin{align}
    \exp \left[ i q \oint_{C_1^{0k}} ( B^i - B^j )\right] \simeq \exp \left[ i q \oint_{C_1^{0k}} ( dx^0 \hat{A}_0^{k(ij)} + dx^k \hat{A}^{ij}) \right] \,.
\end{align}
Then we can derive the field correspondences
\begin{subequations}
\begin{align}
    B^i_0 - B^j_0 &\simeq \hat{A}_0^{k(ij)} \,, \quad 
 (i,j,k) = (1,2,3), (2,3,1), (3,1,2) \,, \label{3corr4} \\
    B^i_k - B^j_k &\simeq \hat{A}^{ij} \,, \quad (i,j,k) = (1,2,3), (2,3,1), (3,1,2) \,. \label{3corr5}
\end{align}
\end{subequations}
The gauge transformations by $\mu$ cancel out in the left-hand sides and the gauge transformations by $\beta^k$ do not appear. Note that $B^2_0 - B^1_0 \simeq -\hat{A}_0^{3(21)}$, $B^2_3 - B^1_3 \simeq -\hat{A}^{21}$ and so on. These correspondences are consistent with the conditions $\hat{A}_0^{1(23)} + \hat{A}_0^{2(31)} + \hat{A}_0^{3(12)} = 0$. From (\ref{3fgauge2}), (\ref{3egauge3}) and (\ref{3egauge4}), the gauge transformations are
\begin{subequations}
\begin{align}
    B^i_0 - B^j_0 &\rightarrow B^i_0 - B^j_0 +  \partial_0 ( \lambda^i - \lambda^j) \,, \\
    B^i_k - B^j_k &\rightarrow B^i_k - B^j_k + \partial_k ( \lambda^i - \lambda^j) \,, \\
    \hat{A}_0^{k(ij)} &\rightarrow  \hat{A}_0^{k(ij)} 
 + \partial_0 \hat{\alpha}^{k(ij)} \,, \\
    \hat{A}^{ij} &\rightarrow \hat{A}^{ij} + \partial_k \hat{\alpha}^{k(ij)} \,,
\end{align}    
\end{subequations}
where $(i,j,k) = (1,2,3), (2,3,1), (3,1,2)$. Then we obtain the gauge parameter correspondences
\begin{align}
     \lambda^i - \lambda^j &\simeq  \hat{\alpha}^{k(ij)} \,, \quad (i,j,k) = (1,2,3), (2,3,1), (3,1,2) \,.
\end{align}    
Moreover, the gauge transformations of $\lambda^k$ and $\hat{\alpha}^{k(ij)}$ are
\begin{subequations}
\begin{align}
    \lambda^k &\rightarrow \lambda^k +  2\pi m^k + \nu \,, \\
    \hat{\alpha}^{k(ij)} &\rightarrow  \hat{\alpha}^{k(ij)} 
 + 2\pi\tilde{m}^i - 2\pi \tilde{m}^j \,,
\end{align}    
\end{subequations}
which can be matched by
\begin{subequations}
\begin{align}
     m^k &\simeq \tilde{m}^k \,, \quad k=1,2,3 \,.
\end{align}  
\end{subequations}
Again in these correspondences, their singularities and discontinuities are matched. 
Using the correspondences \eqref{3corr4} and \eqref{3corr5}, we find the gauge-invariant operator corresponding to $\tilde{P}^q_{3,1}$ defined in \eqref{3diplin1}:
\begin{equation}
\begin{split}
    P^q_{3,1}[S^3_2]  = \exp \bigg[ i q \oint_{S^3_2} \left\{ dx^0 dx^3 \partial_3 (B_0^2 -B_0^3) + dx^1 dx^3 \partial_3 (B_1^2 -B_1^3) \right.  \\
     \left. - dx^2 dx^3 \left( \partial_3 (B_2^3 - B_2^1 ) + \partial_2 ( B_3^1 - B_3^2) \right) \right\} \bigg]  \,.
\end{split}
\end{equation}
Similarly, we also have $P^q_{k,i}$ for $(k,i)=(3,2),(1,2),(1,3),(2,1),(2,3)$.

Under the correspondence \eqref{3corr5}, the cage time-like symmetry operator $T^q[S_2^{123,\text{cube}}]$ defined \eqref{3ftimelike} corresponds to $\tilde{T}^q[S_2^{123,\text{cube}}]$ defined in \eqref{3etimelike1}. Note that on a Hilbert space with fracton defect operators, the $b$ operator $T^q[S_2]$ is a product of $T^q[S_2^{123,\text{cube}}]$ surrounding the defects that are surrounded by $S_2$. In addition, using the correspondences \eqref{3corr3}, we can find the belt time-like symmetry operator in the foliated side corresponding to $\tilde{U}^q_{[12]3}\left[ S_2^{3,\text{belt}}\right]$ defined in \eqref{3etimelike2}:
\begin{equation}
\begin{split}
    U_{[12]3,\text{mod}}^q\left[S_2^{3,\text{belt}}\right] &= \exp \left[ i q \int_{x^2_1}^{x^2_2} \int_{x^3_1}^{x^3_2}  \left(A^3_2(x^1_2) + \partial_2 a_3(x^1_2) - A^3_2 (x^1_1)  - \partial_2 a_3 (x^1_1)\right) dx^2\wedge dx^3  \right] \\
    & \quad   \times \exp \left[- i q \int_{x^1_1}^{x^1_2} \int_{x^3_1}^{x^3_2}  \left(A^3_1(x^2_2)  + \partial_1 a_3(x^2_2) - A^3_1 (x^2_1) - \partial_1 a_3 (x^2_1)\right) dx^1\wedge dx^3  \right]  \,,
\end{split}
\end{equation}
or non-modified one 
\begin{equation}
\begin{split}
U_{[12]3}^q\left[S_2^{3,\text{belt}}\right] &= \exp \left[ i q \int_{x^2_1}^{x^2_2} \int_{x^3_1}^{x^3_2}  \left(A^3_2(x^1_2)   - A^3_2 (x^1_1) \right) dx^2\wedge dx^3  \right] \\
    & \quad   \times \exp \left[- i q \int_{x^1_1}^{x^1_2} \int_{x^3_1}^{x^3_2}  \left(A^3_1(x^2_2)   - A^3_1 (x^2_1) \right) dx^1\wedge dx^3  \right]  \,.
    \end{split}
\end{equation}
Similarly, we can find the other belt operators $U_{[31]2}^q\left[S_2^{2,\text{belt}}\right]$ and $U_{[23]1}^q\left[S_2^{1,\text{belt}}\right]$.

There are other gauge-invariant operators, e.g.\ the one describes the creation of a quadrupole of fractons \cite{Seiberg:2020cxy,Seiberg:2020wsg} in the exotic side, and one can easily map them to the foliated side using the correspondences of the gauge fields.

%% file: conclusion.tex
\section{Conclusion}

In this paper, we have discussed the duality between the foliated $BF$ theory and the exotic $BF$ theory in $2+1$ dimensions and $3+1$ dimensions, and derived the explicit correspondences of the gauge fields and parameters by matching the gauge-invariant operators. The correspondences include the bulk field in the FQFT, and the singularities and discontinuities of the fields and the parameters are also consistent. 
The duality includes the correspondences of the time-like symmetries \cite{Gorantla:2022eem} in both sides.

One of the future directions is to consider the mixed 't Hooft anomalies \cite{tHooft:1979rat} of the subsystem symmetries. The 't Hooft anomalies and the anomaly inflow \cite{Callan:1984sa} of the subsystem symmetries are studied \cite{Yamaguchi:2021xeq,Burnell:2021reh}, and the corresponding symmetry protected topological (SPT) phases are called subsystem SPT (SSPT) phases \cite{You:2018oai,Devakul:2018fhz,Burnell:2021reh}. The foliated and exotic $BF$ theory in this paper also have the mixed anomalies, and it would be interesting to consider how the duality incorporate the SSPT phases. 

Another direction is to consider relations between gapless foliated and exotic QFTs such as the foliated scalar \cite{Hsin:2021mjn,Geng:2021cmq} and the $\phi$ theory \cite{Seiberg:2020bhn,Gorantla:2020xap}. Although in gapless theories the foliated and exotic theories will not represent the same physics, as pointed out in \cite{Hsin:2021mjn}, they might be connected by a renormalization group flow.

In this work, we have considered the foliated $BF$ theory in the flat foliations. It is not understood so far what exotic tensor gauge theory corresponds to the general foliated $BF$ theory \eqref{flagrangian}. 
If there are foliated-exotic dualities for more general classes of foliated theories, e.g.\ the ones studied in \cite{Slagle:2018swq,Slagle:2020ugk,Hsin:2021mjn}, it would provide a more general construction of exotic QFTs.

%% file: appendixa.tex
\appendix
\section{Electric-Magnetic Dual Description in 2+1 Dimensions} \label{appendixa}

The ordinary $BF$ theory in $1+1$ dimensions is dual to the $\Z_N$ gauge theory realized as Higgsing of a charge-$N$ scalar field coupled to a $U(1)$ gauge field \cite{Maldacena:2001ss,Banks:2010zn,Kapustin:2014gua,Gukov:2013zka}, which is the electric-magnetic duality.
Similarly the foliated $BF$ theory described in Section \ref{2+1 fBF} and the exotic $BF$ theory described Section \ref{2+1 eBF} can be written as such a dual description. These electric-magnetic dual theories are directly foliated-exotic dual to each other. 

\subsection{Foliated Gauge Theory}

In the foliated $BF$ theory, the term of a stack of $1+1$d $BF$ theories can be written as the form of a $\Z_N$ gauge theory realized as Higgsing. The Lagrangian is 
\begin{align}
\L'_f = \sum^{n_f}_{k=1} \left[ -\frac{i}{2\pi} U^k \wedge (d\Phi^k - N A^k ) \wedge dx^k + \frac{i N}{2\pi} b \wedge A^k \wedge dx^k \right]   + \frac{i N}{2\pi} b \wedge d a\,, \label{fscalar}
\end{align}
where $\Phi^k$ is a compact scalar field and $U^k$ is a one-form field, which is a Lagrangian multiplier. $\Phi^k$ can have delta function singularities in the $x^k$ direction, and $U^k$ can have zero-form step function discontinuities 
 and one-form delta function singularities in the $x^k$ direction. %
The gauge transformation is
\begin{subequations}
\begin{align}
     \Phi^k &\rightarrow  \Phi^k  + N \zeta^k + 2\pi \partial_k t^k \,, 
\end{align}
\end{subequations}
where $t^k$ is a $x^k$-dependent function valued in integers that can have step function discontinuities in the $x^k$ direction, and $\zeta^k$ is defined in Section \ref{ffield} and Section \ref{21fbf}.
The equation of motion derived by integrating the Lagrangian multiplier out is
\begin{align}
    (d \Phi^k - N A^k)\wedge dx^k = 0\,. \label{phikeom}
\end{align}
Then the strip operators (\ref{fstrip}) can be written as
\begin{align}
    W_k^q[S^k_2] = \exp \left[ i \frac{q}{N} \oint_{S_2^k} d \Phi^k \wedge dx^k \right] \,, \quad k=1,2\,. \label{fstripscalar}
\end{align}

We can dualize this theory to the foliated $BF$ theory. By integrating $\Phi^k$ out, we can derive the equation of motion
\begin{align}
    \frac{N}{2\pi} d U^k \wedge dx^k = 0\,.
\end{align}
Solving this equation locally, we can derive $U^k \wedge dx^k = dB^k \wedge dx^k$, where $B^k$ is the foliated $B$-type zero-form field defined in Section \ref{ffield}. Then the Lagrangian becomes
\begin{align}
    \L'_f \rightarrow \sum^{n_f}_{k=1} \left[ \frac{i N}{2\pi} dB^k \wedge A^k  \wedge dx^k + \frac{i N}{2\pi} b \wedge A^k \wedge dx^k \right]   + \frac{i N}{2\pi} b \wedge d a\,,
\end{align}
which is equal to the foliated $BF$ Lagrangian (\ref{21folilag}).

\subsection{Exotic Gauge Theory}

The exotic $BF$ theory also has the form of the $\Z_N$ tensor gauge theory realized as Higgsing of a charge-$N$ scalar field coupled to a $U(1)$ tensor gauge field \cite{Seiberg:2020bhn}.
This $\Z_N$ tensor theory is directly dual to the foliated $\Z_N$ gauge theory (\ref{fscalar}). The Lagrangian is
\begin{align}
\L'_e = \frac{i}{2\pi} \hat{E}^{12} ( \partial_1 \partial_2 \phi - NA_{12}) + \frac{i}{2\pi} \hat{B} (\partial_0 \phi - N A_0) \,,
\end{align}
where $(\hat{E}^{12}, \hat{B})$ in the representation $(\bm{1}_2,\bm{1}_0)$ is a Lagrangian multiplier and $\phi$ in the representation $\bm{1}_0$ is a compact scalar. The field $\phi$ can have configurations that are the sum of the terms proportional to step functions in the $x^1$ and $x^2$ directions. The gauge transformation is
\begin{align}
    \phi \rightarrow \phi + N\alpha + 2\pi \tilde{t}^{\,1}  + 2\pi \tilde{t}^{\,2} \,,
\end{align}
where $\tilde{t}^{\,k}$ are $x^k$-dependent functions valued in integers that can have step function discontinuities in the $x^k$ direction, and $\alpha$ is defined in Section \ref{tensor}.

The equations of motion derived by integrating the Lagrangian multiplier out are
\begin{subequations}
\begin{align}
    \partial_1 \partial_2 \phi - NA_{12} &= 0 \,, \label{phieom1} \\
    \partial_0 \phi - N A_0 &= 0\,. \label{phieom2} 
\end{align}
\end{subequations}
Then the strip operators (\ref{estrip1}) and (\ref{estrip2}) can be written as
\begin{subequations}
\begin{align}
    \tilde{W}_1^q\left[ S^1_2 \right] = \exp \left[ i \frac{q}{N} \oint_{S^1_2} ( dx^0 dx^1 \partial_1 \partial_0 \phi + dx^2 dx^1 \partial_1 \partial_2 \phi )  \right] \,, \label{escalarstrip1}\\
    \tilde{W}_2^q\left[ S^2_2  \right] = \exp \left[ i \frac{q}{N} \oint_{S^2_2} ( dx^0 dx^2 \partial_2 \partial_0 \phi + dx^1 dx^2 \partial_1 \partial_2 \phi )  \right]\,. \label{escalarstrip2}
\end{align}
\end{subequations}

We can dualize this theory to the exotic $BF$ theory. Integrating $\phi$ out, we can derive the equation of motion
\begin{align}
    \partial_1 \partial_2 \hat{E}^{12} - \partial_0 \hat{B} = 0\,.
\end{align}
Solving this equation locally, we can derive $\hat{E}^{12} = \partial_0 \phi^{12}$ and $\hat{B} = \partial_1 \partial_2 \phi^{12}$. Then the Lagrangian becomes
\begin{align}
    \L'_e \rightarrow \frac{i N}{2\pi} \phi^{12}  \partial_0  A_{12} - \frac{i N}{2\pi} \phi^{12} \partial_1 \partial_2 A_0 \,,
\end{align}
which is equal to the exotic $BF$ Lagrangian (\ref{elagrangian}).

\subsection{Correspondences}

In the dual theories, from the strip operators (\ref{fstripscalar}), (\ref{escalarstrip1}), (\ref{escalarstrip2}), and the modification of $W^q_k[S^k_2]$ in \eqref{modstrip}, we identify
\begin{subequations}
\begin{align}
    \exp \left[ i \frac{q}{N} \oint_{S^1_2} \left( d \Phi^1  \wedge dx^1 +  N d(a_1 dx^1) \right) \right]  &\simeq 
    \exp \left[ i \frac{q}{N} \oint_{S^1_2} ( dx^0 dx^1 \partial_1 \partial_0 \phi + dx^2 dx^1 \partial_1 \partial_2 \phi )  \right] \,, \\
    \exp \left[ i \frac{q}{N} \oint_{S^2_2} \left( d \Phi^2  \wedge dx^2 + N d(a_2 dx^2) \right) \right]  &\simeq 
    \exp \left[ i \frac{q}{N} \oint_{S^2_2} ( dx^0 dx^2 \partial_2 \partial_0 \phi + dx^1 dx^2 \partial_1 \partial_2 \phi )  \right] \,,
\end{align}
\end{subequations}
which lead to the correspondences of the scalar fields
\begin{subequations}
\begin{align}
    \Phi^1 + N  a_1 &\simeq \partial_1 \phi \,, \\ 
    \Phi^2 + N  a_2 &\simeq  \partial_2 \phi \,.
\end{align}
\end{subequations}
The gauge transformations of $\Phi^k$, $a$ and $\phi$ are
\begin{subequations}
\begin{align}
    \Phi^k &\rightarrow  \Phi^k + N\zeta^k + 2\pi \partial_k t^k  \,, \\ 
    a_k &\rightarrow  a_k + \partial_k \lambda - \zeta^k \,, \\
    \phi &\rightarrow  \phi + N \alpha + 2\pi \tilde{t}^{\,1} + 2\pi \tilde{t}^{\,2} \,. 
\end{align}
\end{subequations}
The gauge transformation by $\zeta^k$ cancel out in the left-hand sides. Then we derive the gauge parameter correspondences
\begin{subequations}
\begin{align}
    \lambda &\simeq \alpha \,, \\ 
    t^k &\simeq  \tilde{t}^{\,k} \,.
\end{align}
\end{subequations}
Again the discontinuities are matched.